\documentclass[journal]{vgtc}                     

\newcommand{\revise}[1]{\textcolor{black}{#1}}

\newcommand{\dhkim}[1]{
\textcolor{black}{#1}}

\onlineid{1484}

\vgtccategory{Research}

\vgtcpapertype{please specify}

\title{
Evaluating Preattentive Features for Detecting Changes \\ in Virtual Environments
}

\author{%
  \authororcid{DongHoon Kim}{0000-0002-9969-8394},
    and 
  \authororcid{Isaac Cho}{0000-0003-1582-8428}, \textit{Member, IEEE }
}

\authorfooter{
  \item
  	DongHoon Kim is with Utah State University.
  	E-mail: donghoon.kim@usu.edu
  \item
  	Isaac Cho is with Utah State University (corresponding author).
  	E-mail: isaac.cho@usu.edu.
}

\abstract{%
Visual perception plays a critical role in detecting changes within immersive Virtual Reality (VR) environments. However, as visual complexity increases, perceptual performance declines, making it more difficult to detect changes quickly and accurately.
This study examines how visual features, known for facilitating preattentive processing, impact a change detection task in immersive 3D environments, with a focus on visual complexity, object attributes, and spatial proximity.
Our results demonstrate that preattentive processing enhances change detection, particularly when the altered object is spatially isolated and not perceptually grouped with similar surrounding objects. Changes to isolated objects were detected more reliably, suggesting that perceptual isolation reduces cognitive load and draws more attention. Conversely, when a changed object was surrounded by visually similar elements, participants were less likely to detect the change, indicating that perceptual grouping hinders individual object recognition in complex scenes.
These results provide guidelines for designing VR applications that strategically utilize spatial isolation and visual features to improve the user experience.
}


\teaser{
  \centering
  \includegraphics[width=\linewidth, alt={A view of a city with buildings peeking out of the clouds.}]{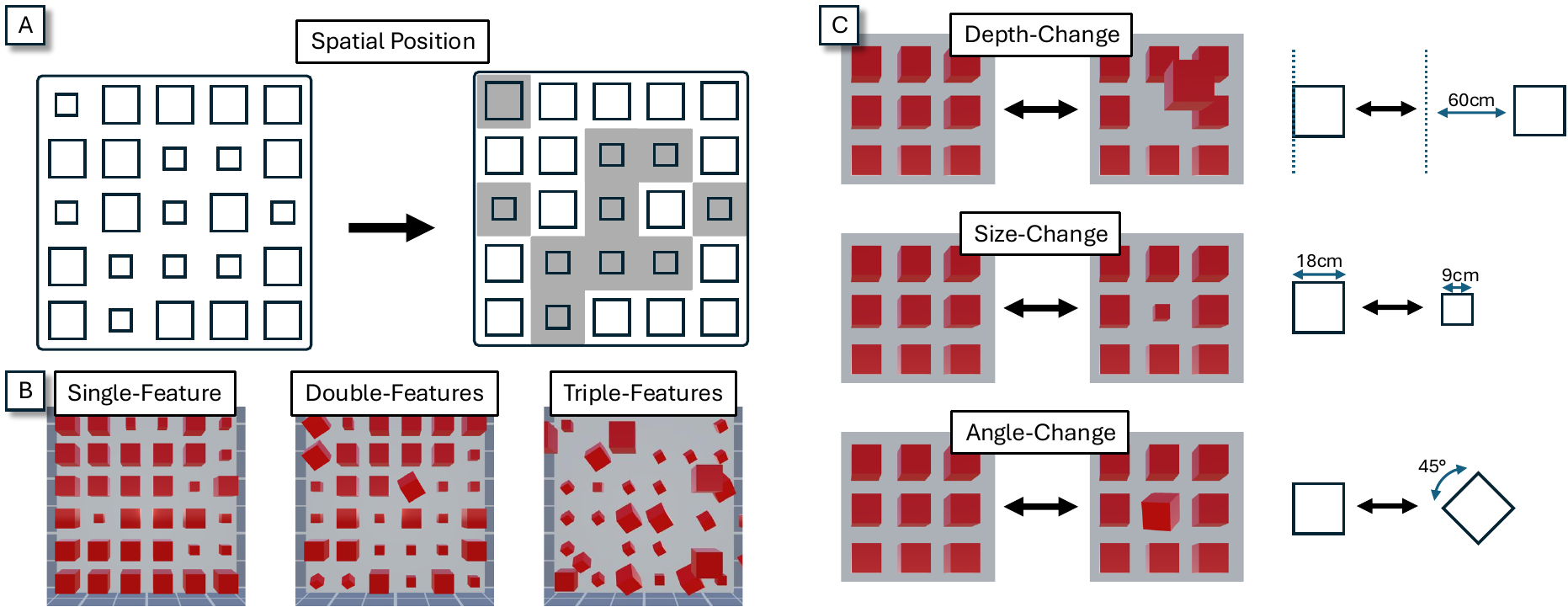}
  \caption{%
Our study evaluates the effects of preattentive features, visual complexity, and spatial position on change detection performance in a VR environment. Specifically, we examine how (A) the spatial position of the target object (Isolated vs. Grouped), (B) visual complexity determined by the number of features (Single, Double, Triple), and (C) changed features (Depth, Size, Angle) influence users’ ability to detect visual changes.
  }
  \label{fig:teaser}
}

\graphicspath{{figs/}{figures/}{pictures/}{images/}{./}} 
\usepackage{tabu}                      
\usepackage{booktabs}                  
\usepackage{lipsum}                    
\usepackage{mwe}                       

\usepackage{mathptmx}                  

\begin{document}


\firstsection{Introduction}

\maketitle

Human perception is fundamental in Virtual Reality (VR), where vision is typically the dominant sensory modality.
VR systems heavily rely on visual stimuli to enhance the sense of presence, while other sensory inputs also contribute to immersion.
Given that visual processing accounts for a significant portion of human sensory input, misperceptions in VR can lead to errors in spatial awareness, interaction accuracy, and environmental interpretation.

Accurately perceiving and interpreting multiple objects within a limited time window is inherently challenging due to the constraints of visual short-term memory (VSTM)
which can compromise a user's ability to process and retain information when presented with several objects simultaneously~\cite{buss2018visual}.
However, certain objects can be detected instantly due to the \textbf{preattentive features}, distinct visual properties, such as shape, color, or orientation.
These features enhance perceptual efficiency by enabling rapid, automatic detection, reducing the need for focused attention.
Preattentive processing allows certain objects to stand out due to their contrast with the surrounding environment ~\cite{ware2019information}.
This process occurs automatically, enabling rapid detection regardless of the type of difference or number of surrounding objects or the complexity of the scene~\cite{wolfe1992role, nagy1990visual, enns1990three}.
Because of its speed and efficiency, preattentive processing helps encode objects into memory more effectively and facilitates rapid target identification, even in visually complex environments.

In data visualization, the advantages of preattentive processing are particularly important~\cite{ware2019information}.
Visualization systems that leverage preattentively processable visual stimuli can effectively represent complex data using different preattentive features, such as size, length, shape, or orientation. For example, a large symbol near other small symbols naturally draws attention, making it the first to be perceived. Similarly, a horizontal bar among vertical bars or a red marker among yellow markers can be quickly and effortlessly identified.
Much research has explored the mechanisms and characteristics of preattentive processing, primarily concentrating on 2D display environments~\cite{barrera2023preattentive}.
However, due to the perceptual differences between 2D and 3D environments, it is challenging to apply prior research results from 2D environments directly to immersive environments~\cite{snow2014real,korisky2021dimensions,sagehorn2024comparative}.
Previous studies on change detection in immersive environments, in contrast, have mainly focused on non-preattentive features \cite{steinicke2011change,senel2023imperceptible, martin2023study}, leaving open the question of how preattentive features operate in immersive settings. To address this gap, we systematically explore how combinations of preattentive features affect change detection in an immersive VR environment.  To guide our investigation, we focus on the following research questions:

\begin{enumerate}
     \item [\textbf{RQ1}:] 
     How do preattentive features influence change detection performance in VR, and how do their effects vary across different feature types?

    \item [\textbf{RQ2:}] How does increasing visual complexity, operationalized by the number of \revise{pre}attentive features, affect change detection performance and perceived workload?

\item [\textbf{RQ3}:]
 How does the spatial separation of a changed object affect change detection performance in a VR environment?
 \end{enumerate}

To address these questions, we conducted a change detection experiment in which participants identified a changed object within a VR scene containing multiple non-changing distractors. Our results show that preattentive features facilitate rapid change detection in visually simple scenes, but their effectiveness declines as visual complexity increases. Among the examined features, depth changes were detected more quickly and accurately than size or angle changes. Moreover, spatial isolation consistently enhanced detection performance, especially under complex conditions, underscoring its value for directing attention and reducing cognitive load in VR environments.
\section{Related Works}
\subsection{Preattentive Features and Visual Search}
Preattentive features are basic visual properties that allow the human visual system to rapidly detect certain stimuli without conscious effort~\cite{treisman1980feature, wolfe2019preattentive}. These features make objects stand out based on their contrast with surrounding elements, enabling efficient guidance of attention during visual search. For example, when an object differs in size from its surroundings, attention is automatically drawn to the target~\cite{treisman1980feature, wolfe2019preattentive}. Similarly, variations in orientation~\cite{wolfe1992role}, color~\cite{nagy1990visual}, and depth in stereoscopic environments~\cite{enns1990three} enhance detectability through preattentive features~\cite{healey1996high}. Each type of feature has a distinct level of salience. A study on visual attention~\cite{hadnett2019effect} showed that, although task type can influence attention, feature characteristics such as saliency, color, intensity, and orientation also play a critical role. When comparing a color and shape, color changes are perceived more saliently than changes in shape when visual information is presented for a short time~\cite{barreiros2016pre}.

The effectiveness of attention guidance largely depends on the type and magnitude of feature contrast between the target and its surroundings~\cite{cheal1992attention}. As a result, visual search performance improves when the target object is highly distinct from the background~\cite{healey2011attention}. When objects differ in only one preattentive feature (e.g., color), visual stimuli are typically processed in parallel, allowing for rapid detection~\cite{treisman1980feature, treisman1982perceptual, treisman1985preattentive}. However, when multiple feature differences are present, a condition known as conjunction search~\cite{krekhov2019deadeye}, the task typically requires serial processing, which increases cognitive effort and slows detection~\cite{treisman1988feature}. For example, combining spatial and color differences or pairing features such as color, motion, size, and orientation, results in slower search performance due to the need for feature-by-feature comparison.

The degree of performance decline in conjunction search also varies across feature types. Hulleman et al.~\cite{hulleman2020medium} reported that adding orientation differences worsens search performance more than adding color differences~\cite{wolfe2020forty}, as orientation variations increase cognitive load. This pattern persists even when three or more features vary simultaneously~\cite{wolfe1989guided}. Interestingly, depth has been shown to function differently from other features. Nakayama and Silverman~\cite{nakayama1986serial} demonstrated that depth differences can preserve parallel processing even when combined with other feature variations, such as color or motion. Studies on visual working memory and object tracking further suggest that depth improves visual perception by reducing interference from nearby objects. For example, depth separation facilitates movement tracking by preventing objects at different depths from being perceived as a single group~\cite{viswanathan2002dynamics}. Similarly, Xu and Nakayama~\cite{xu2007visual} found that placing objects on separate 3D surfaces enhanced change detection performance, indicating that depth separation supports more efficient visual processing.

While prior research has shown the benefits of preattentive features, most studies have focused on 2D displays, leaving their effects in immersive VR environments underexplored. Therefore, this study investigates how features such as depth, size, and orientation influence change detection in VR.

\subsection{Change Detection}
Visual short-term memory (VSTM)\revise{-}also known as working memory~\cite{cowan2008differences}\revise{-}allows humans to temporarily retain and process visual information, even during brief interruptions such as eye movements~\cite{peterson2013gestalt}. While VSTM enables rapid storage of visual content, its capacity is highly limited, restricting the number of objects it can hold at a given time.

Prior research on change detection tasks~\cite{alvarez2004capacity, awh2007visual} and neuroimaging studies using functional magnetic resonance imaging (fMRI)~\cite{xu2006dissociable} indicate that VSTM can store approximately four objects simultaneously. When the number of visual stimuli exceeds this capacity, observers struggle to encode and retain information, often resulting in failures to detect changes~\cite{simons2005change, cowan2005capacity, moriya2019visual}. Simons et al.~\cite{simons1997change} demonstrated that detection failures can occur even when a change happens directly within the observer’s field of view. For instance, participants often fail to notice changes in an object's color or shape, or even a change in a person's identity during interaction~\cite{simons1998failure, levin2002memory, varakin2007comparison}. These findings highlight the limitations of VSTM in dynamic visual environments and its critical role in change detection performance.

Moreover, change detection performance varies depending on the type of object attribute involved. According to the working memory model, intrinsic object properties, such as shape and color, are processed differently from spatial information~\cite{h1997inner}. Specifically, the visuospatial sketchpad model suggests that visual information is processed through two subcomponents: the visual cache, which stores object-specific features (e.g., color, shape), and the inner scribe, which handles spatial properties (e.g., location, movement). Although these subcomponents develop at different rates across human ages, it remains unclear whether their performance differs for adults in complex environments.

In immersive VR environments, rich depth cues and head-tracked rendering can enhance visual perception. However, successful change detection still requires efficient encoding and maintenance of visual information within VSTM~\cite{martin2023study, kim2023comparative, steinicke2010change}. Previous change detection studies were either conducted without systematic environmental control ~\cite{martin2023study},  or used changeable objects placed at random~\cite{kim2023comparative}, limiting the ability to isolate specific influencing factors. Moreover, limited attention has been given to how preattentive features affect change detection performance, particularly in visually complex VR environments. Our study addresses this gap by systematically investigating feature-specific effects on change detection performance in VR. 

\subsection{Perception in VR}
Immersion is a key characteristic of VR, enabling users to feel present and engaged within a virtual environment. A primary technique for achieving immersion is head-tracked rendering, which dynamically updates the visual scene based on the user’s head position and orientation~\cite{bowman2007virtual}. This technique enhances users' sense of presence by creating a responsive and interactive visual experience. However, despite this immersive experience, user perception in VR often differs from perception in the real world. Prior research has identified several types of perceptual distortions that arise in VR environments~\cite{bansal2019movement, mullen2021time}.

Spatial perception is also subject to distortion in VR. Numerous studies have found that users tend to underestimate distances to objects in VR compared to the real world~\cite{plumert2005distance}. This underestimation is often attributed to a lack of reliable visual cues or the presentation of visual information in a distorted form. Object size perception can also be inaccurate in VR environments~\cite{rzepka2023familiar}, raising further concerns about users’ ability to interpret spatial relationships. Depth perception, in particular, has been a critical focus in VR research because it directly supports spatial understanding, object localization, and interaction performance. Prior work has shown that enhancing visual fidelity~\cite{drascic1996perceptual} and providing multiple depth cues~\cite{surdick1997perception} can improve depth perception in virtual environments. Depth cues play a critical role in enhancing visual search and object recognition, enabling faster and more accurate detection of depth-based changes compared to other feature changes~\cite{drascic1996perceptual, surdick1997perception, cho2012evaluating}. Head-tracking techniques further support depth perception by providing motion parallax and dynamic spatial cues~\cite{cho2012evaluating, creem2005influence}. However, depth perception in VR remains vulnerable to technical limitations, such as the vergence-accommodation conflict in head-mounted displays~\cite{vienne2020depth}.

Prior research also indicates that limited VSTM capacity disrupts change detection. Nevertheless, preattentive features may help mitigate this limitation by facilitating rapid, low-level encoding. Yet, because these principles have been predominantly established in 2D environments, their applicability in immersive VR contexts remains insufficiently explored.

\section{Experiment Design}

\begin{figure}
\includegraphics[angle=0, width=.485\textwidth]{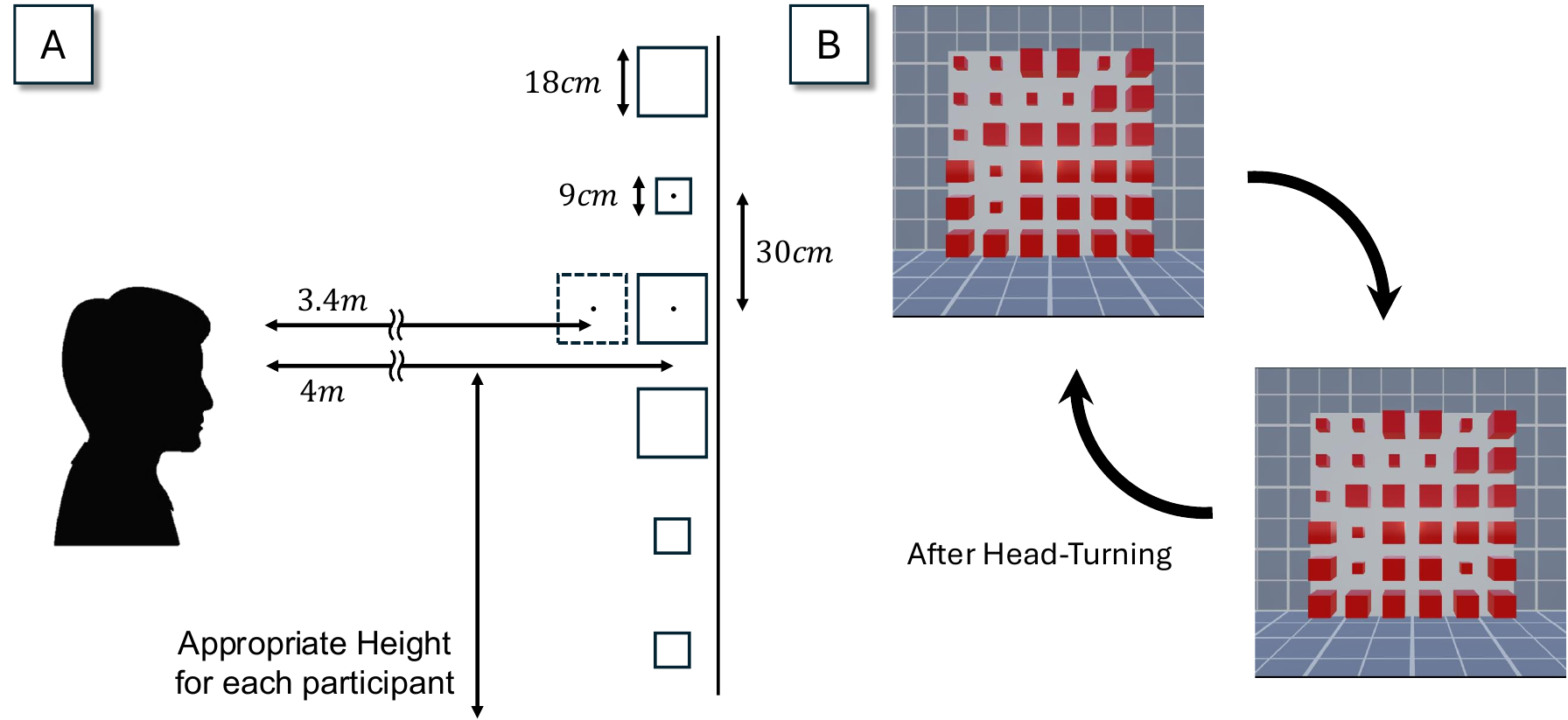}
\caption{
A) Illustration of the experimental environment. A 6 $\times$ 6 grid-patterned board is placed \revise{3.4 m or 4 m (depending on the depth condition)} in front of each participant, with its height (at the center of the board) adjusted to match each participant's eye level.
B) One of the objects is changed after a participant turns their head. Participants have to find and select one that has changed.
}
\vspace{-.5cm}
\label{figure:experiment}
\end{figure}

\subsection{Virtual Environment}
The experiment is conducted in a monotonously designed 5m $\times$ 5m $\times$ 5m VR environment (Fig.~\ref{figure:experiment}) designed to minimize extraneous visual stimuli. Participants view a 6 $\times$ 6 grid of uniformly shaped red cubes. While all objects share the same color and shape, their preattentive features (\textit{Depth}, \textit{Size}, and \textit{Angle}) are assigned based on the experimental conditions.

The experiment manipulated three independent variables: \textbf{Feature Type} (\textit{Depth}, \textit{Size}, \textit{Angle}), \textbf{Number of Features} (\textit{Single, Double, Triple}), and \textbf{Separation} (\textit{Isolated} vs. \textit{Grouped}).
\dhkim{
The \textbf{Feature Type} condition regulates a method for changing one of the objects after a participant's sight is disrupted.
The \textbf{Number of Features} condition determines how many visual features are used for grouping, although the change always occurs in only one feature specified by the \textbf{Feature Type}.
The \textbf{Separation} condition controls the spatial arrangement of the target object, which is either positioned in isolation or surrounded by similar objects.
}

The distance between the centers of adjacent cubes is fixed at 30 cm. Participants are positioned 4m away from the reference plane and perform the change detection task at this location (Fig.~\ref{figure:experiment}A).
\dhkim{
To simulate natural attentional shifts, a visual change was triggered only when the participant's head turn caused the display grid to move completely out of the field of view. This trigger mechanism is based on common user behaviors in VR, such as scanning the environment or navigating.
}
To ensure participants fully turn their head, a gray cylinder placed on both the left and right sides changes its color to green when the target object is completely outside their view.
Participants are required to turn their head back and forth multiple times, exploring the scene until they detect the change.
\revise{Each head turn alternated the target object between its original and modified states.}
For example, as shown in Fig.~\ref {figure:experiment}B, the target object in the bottom right corner changes its size \revise{between} small and large.

\subsection{Experiment Conditions}
\subsubsection{Feature Type}
Based on prior work showing that performance varies by visual feature type~\cite{nakayama1986serial}, we included \textbf{Feature Type} as an independent variable to investigate how different preattentive features affect change detection.
Among various preattentive features, \textit{Size} and \textit{Angle} offer the most precise quantitative control compared to other primary features such as texture, color, or shape~\cite{roth2017visual}.
Also, according to Nakayama and Silverman~\cite{nakayama1986serial}, \revise{simulated geometric} \textit{Depth} differences are perceptually distinguishable, making \textit{Depth} a suitable feature to examine in an immersive environment.
Therefore, we selected \textit{Depth}, \textit{Size}, and \textit{Angle} as the \textbf{Feature Type} conditions in this study. \revise{While \textit{Depth} does not refer to optical focal depth but to virtual object distance defined in the 3D scene}, comparing it with \textit{Size} and \textit{Angle} provides valuable insights into the role of different preattentive features in change detection, particularly as VR environments provide multiple cues to enhance depth perception such as stereopsis, and head-tracking \cite{cho2012evaluating}.

In the \textit{Size} condition, the target object's size alternates between small (9cm $\times$ 9cm $\times$ 9cm) and large (18cm $\times$ 18cm $\times$ 18cm).
In the \textit{Angle} condition, the object's orientation changes from the the front view (x: 0$^{\circ}$, y: 0$^{\circ}$, z: 0$^{\circ}$) to a corner view (x: 45$^{\circ}$, y: 45$^{\circ}$, z: 45$^{\circ}$) and vice versa.
\revise{In the \textit{Depth} condition, the object's position shifts between close (3.4 m) and far (4 m) from the participant, as shown in Fig. ~\ref{figure:experiment}A.}
Each parameter was determined through an informal pilot study to ensure that visual changes were perceptually salient yet natural within the VR environment.
\revise{Based on informal, author-conducted testing aimed at optimizing stimulus discriminability and user comfort, the following parameter values were selected:}
The rotation angle was set to 45\revise{$^{\circ}$} along each axis (x, y, z) to maximize differentiation without causing discomfort or distraction. Object size was reduced by $50\%$ in length on each axis to maintain visibility and prevent misinterpretation as disappearance. Depth differences were adjusted to avoid occlusion and ensure all objects remained within the participant’s field of view. The spatial layout was designed to minimize overlap and isolate the visual effect of each manipulated feature, accounting for both perceptual salience and VR display constraints.

\subsubsection{Number of Features}
When similarity forms only two groups based on a single preattentive feature, these groups are easily perceived as binary and distinct. However, it is well established that preattentive processing does not operate reliably when multiple features coexist\revise{\cite{wolfe2020forty}}. In particular, previous studies suggest that preattentive features may still support grouping with two features, although the task becomes difficult. For example, two objects may belong to the same group based on \textit{Size} similarity, but \textit{Angle} similarity may divide them into separate groups. In contrast, when three or more features are involved, preattentive processing is known to fail, and grouping no longer occurs automatically \cite{munzner2014visualization}.
To examine this effect, we defined \textbf{Number of Features} as an independent variable to examine how multiple preattentive features interact during change detection. As illustrated in Fig. \ref{fig:teaser}B, the \textit{Single}-feature condition presents objects that differ in only one feature. The \textit{Double}-feature condition involves two feature differences, while the \textit{Triple}-feature condition involves three feature differences between objects. Importantly, although multiple features are used for grouping, the change always occurs in \textbf{only one feature} (i.e., one of \textit{Angle}, \textit{Size}, and \textit{Depth}). In addition, because there are two possible combinations for each feature type condition in the \textit{Double}-feature condition (when a size is changed, the possible \textit{Double}-feature conditions are \textit{Size}-\textit{Angle} and \textit{Size}-\textit{Depth}), the total combination of \textbf{Number of Features} is 4. As the \textbf{Number of Features} increases from one to three, the visual stimulus becomes more complex, potentially affecting preattentive processing and change detection performance.

\subsubsection{Separation}
Similarity among objects facilitates perceptual grouping based on shared characteristics. This grouping by similarity can occur regardless of the spatial proximity between similar objects or the number of such objects present. Consequently, similar objects may form perceptual groups even when they are spatially separated, producing a separation effect.
When an isolated object that shares characteristics with other objects changes, the perceptual group, formed through preattentive processing, appears to dissolve or weaken. In contrast, when an object within a group of similar objects changes, the perceptual grouping remains largely unaffected.

To evaluate change detection performance under these two situations, we set the separation of the change object as an independent variable. In the \textit{Isolated} condition, the changed object was spatially distant from other characteristic-sharing objects, leading preattentive processing to perceive it as a distinct, single-object group. In the \textit{Grouped} condition, the changed object was surrounded by similar objects, and preattentive processing integrated it as a member of the larger group. 

\begin{figure}
\includegraphics[angle=0, width=.47\textwidth]{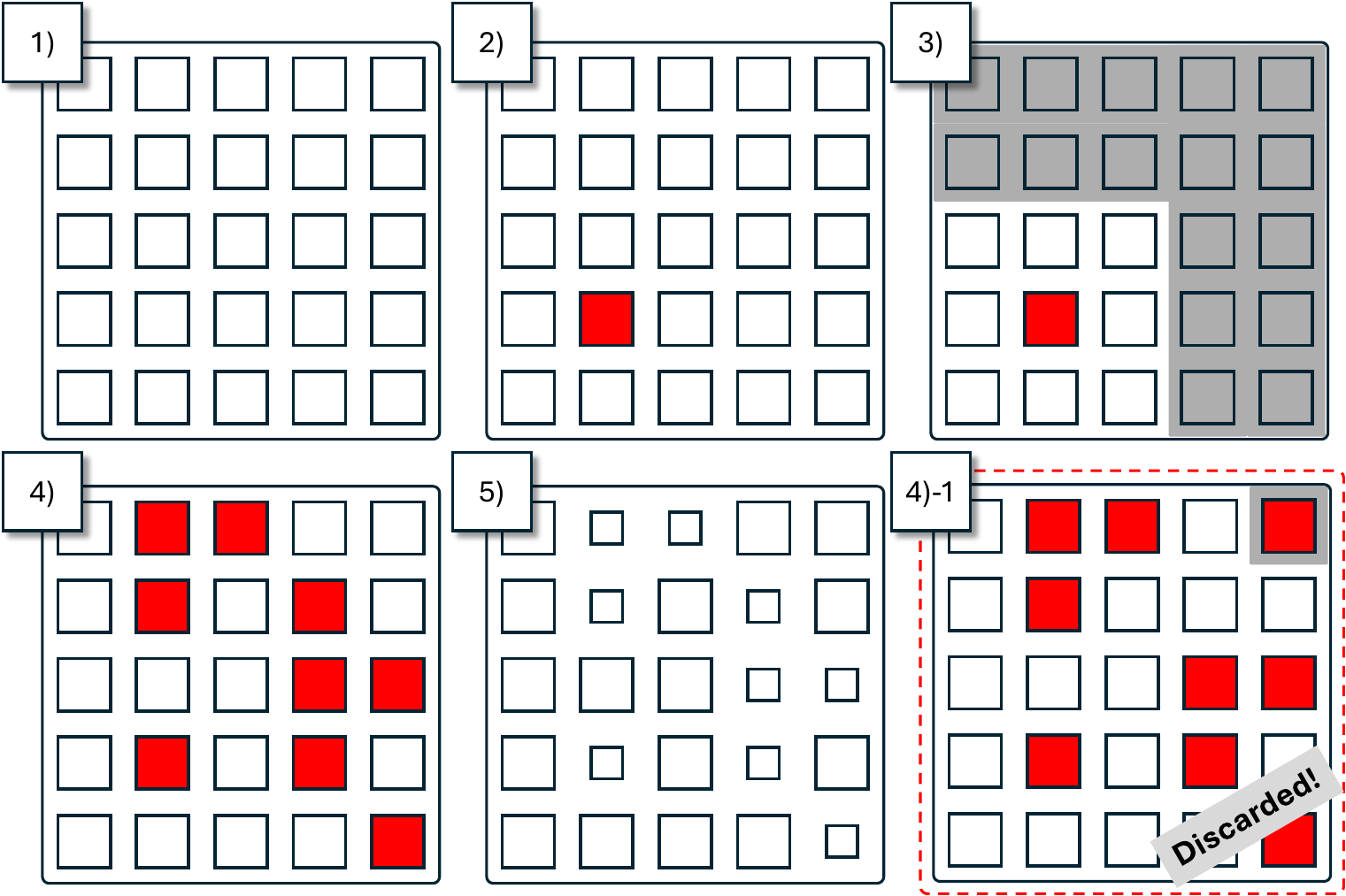}
\caption{
Figure 3: The procedure for generating object composition. Starting with a grid of identical objects (1), one object was selected for the Isolated position (2), and a grouping region was defined away from it (3). Within this region, approximately half of the objects were randomly selected (4) and assigned different visual features from the non-selected objects (5). To ensure grouping, layouts were discarded and regenerated if any selected object lacked a neighboring selected object (4)-1.
}
\label{figure:gridMaking}
\end{figure}

\subsection{Object Composition and Distribution}
The composition and distribution of objects in each task are systematically controlled based on the assigned \textbf{Feature Type} and \textbf{Number of Features} conditions.

In the \textit{Single} condition, all objects share the same visual features except for the feature specified by the Feature Type. For example, if the Feature Type is \textit{Angle}, all cubes are maintained with identical \textit{Size} and \textit{Depth}.
In the \textit{Double} condition, objects vary in one additional feature besides the feature for \textbf{Feature Type}. For example, in the \textit{Depth} change condition, objects can vary in \textit{Depth} and \textit{Size} or \textit{Depth} and \textit{Angle}. However, only \textit{Depth} is altered during the change. In the \textit{Triple} condition, objects vary across all three features (i.e., \textit{Size, Depth, and Angle}), but only one of the features specified by the \textbf{Feature Type} is changed.

We apply the \textit{Isolated} condition only to the feature specified by \textbf{Feature Type}. All other visual features are controlled to form a locally grouped region around the target object. This design prevents participants from directing their attention to the unintended position based on non-changing features. For example, in the \textit{Isolated} and \textit{Size} and \textit{Double} (\textit{Size} and \textit{Angle}) conditions, the target object is visually isolated by having a different \textit{Size} from its surrounding objects. However, it shares the same \textit{Angle} feature with some nearby objects, forming a perceptual group based on \textit{Angle}. This approach prevents the target object from standing out preattentively due to features unrelated to the assigned change.

\revise{The presentation order of the Feature Type and Separation conditions was fully randomized across trials.}
To generate the object layout while satisfying these constraints, one object on the grid is first selected to occupy the position for \textit{Isolated}, that is, isolated based on the assigned \textbf{Feature Type}. For the remaining objects, approximately half are randomly selected (excluding those adjacent to the isolated object) and assigned a shared visual attribute for the non-changing features (e.g., same size, angle, or depth), including the isolated object. This grouping ensures that the target object would not stand out based on non-changing features. The other objects are assigned the contrasting attribute (e.g., large size). If any randomly selected object lacks a valid adjacent object for grouping, the selection process is repeated. Fig.~\ref{figure:gridMaking} illustrates this process.
\revise{Based on the \textbf{Separation} condition, this object was designated as the target in the \textit{Isolated} condition, whereas one of the grouped objects was selected as the target in the \textit{Grouped} condition.}
In the \textit{Double} and \textit{Triple} conditions, this procedure is independently applied to each non-target feature to satisfy the grouping constraints. The layout was regenerated until all composition and distribution constraints were satisfied.

\subsection{Measurements}
To evaluate the effects of the independent variables, we defined four performance measurements: Detection Time, Timeout Rate, Accuracy, and a NASA-Task Load Index (TLX) questionnaire. The first three measurements were recorded during the experimental tasks, while the workload questionnaire was administered after completing the tasks for each Number of Features condition.

\textbf{Detection Time}:
Detection Time measures the duration from the start of a task to the moment a participant selects the object they believe has changed. Faster detection times indicate that the given condition facilitated quicker identification of the feature change. This measurement includes the time spent moving the head, which is required to trigger the feature change by moving the target object out of the participant’s field of view.

\textbf{Timeout Rate}:
Timeout Rate represents the proportion of tasks in which participants failed to detect and select the changed object within the allotted time limit of 60 seconds.
The Timeout Rate also indicates task difficulty. A higher rate suggests that the given condition made change detection substantially more challenging.

The time limit of 60 seconds is determined through an informal pilot test, which showed that users who successfully detected the change typically required less than 45 seconds.

\textbf{Accuracy}:
Accuracy assesses participants’ ability to correctly identify the object that changed its feature under each condition. A lower accuracy indicates that the condition may have disrupted participants’ ability to detect the feature change, either by causing them to overlook the changed object within the time limit or by leading them to mistakenly select an unchanged object. Conversely, a higher accuracy reflects that the condition more effectively supported the detection of feature changes.

\textbf{NASA-TLX:}
NASA-TLX assesses perceived workload across six dimensions: Mental Demand, Physical Demand, Temporal Demand, Performance, Effort, and Frustration ~\cite{hart1988development, hart2006nasa}. Participants rated each dimension using a 7-point Likert scale (1 = Very Low to 7 = Very High; for Performance, 1 = Perfect to 7 = Failure), providing a subjective measure of workload experienced during the task.

\subsection{Hypothesis}
Guided by the research questions and measurements, we formulated the following hypotheses to examine the effects of preattentive features on change detection in immersive VR environments, with respect to visual complexity and feature type.

\begin{enumerate}
    \item [\textbf{H1}] The \textit{Depth} condition is expected to result in faster detection time \textbf{[H1-1]}, lower timeout rate \textbf{[H1-2]}, and higher accuracy \textbf{[H1-3]} compared to the \textit{Size} and \textit{Angle} conditions because VR provides multiple depth cues, such as stereopsis and head-tracking, that facilitate change detection \textbf{(RQ1)}.
    \item [\textbf{H2}] The \textit{Single} condition is expected to result in faster change detection time \textbf{[H2-1]}, lower timeout rates \textbf{[H2-2]}, higher accuracy \textbf{[H2-3]} and compared to the \textit{Double} and \textit{Triple} condition. Similarly, the \textit{Double} condition is expected to lead to faster change detection time \textbf{[H2-4]}, lower timeout rates \textbf{[H2-5]}, and higher accuracy \textbf{[H2-6]} than the \textit{Triple} condition. This is because increasing the number of visual features is likely to disrupt preattentive processing and require more effortful cognitive processing for change detection \cite{munzner2014visualization} \textbf{(RQ2)}.
    \item [\textbf{H3}] The \textit{Isolated} condition is expected to result in faster change detection time \textbf{[H3-1]}, lower timeout rates \textbf{[H3-2]}, and higher accuracy \textbf{[H3-3]} than the \textit{Grouped} condition, as the target object can be more easily perceived through preattentive processing \textbf{(RQ3)}.
    \item [\textbf{H4}] The \textit{Single} condition is expected to have a lower workload \textbf{[H4-1]} than the \textit{Double} and \textit{Triple} conditions. Similarly, the \textit{Double} condition is expected to lead to a lower workload \textbf{[H4-2]} than the \textit{Triple} condition. Similar to \textbf{H2}, this is because increasing the number of visual features requires more effortful cognitive processing for change detection \cite{munzner2014visualization} \textbf{(RQ2)}.
\end{enumerate}


\subsection{Procedures}
\revise{The study protocol was approved by the Institutional Review Board (IRB) of Utah State University (IRB\#14294).} Upon arrival at the study location, participants \revise{were} asked to read and sign a consent form in accordance with the IRB protocol.
They then \revise{completed} a demographic questionnaire, collecting information on age and VR experience.
Following this, the experimenter \revise{explained} the purpose, procedure, and tasks of the study.
Participants \revise{practiced} the experiment tasks, including the required head-turning angle and the method for selecting the changed object.
Each participant \revise{performed} a total of 72 change detection tasks derived from a fully crossed within-subjects design (2 \textbf{Separation} $\times$ 4 \textbf{Number of Features} $\times$ 3 \textbf{Feature Type} $\times$ 3 repetition).
We \revise{counterbalanced} \textbf{Number of Features} across participants using a Latin Square design.
We \revise{randomized} \textbf{Feature Type} and \textbf{Separation} so that participants \revise{could} not anticipate the next trial, and generate object patterns randomly within each trial.
During the experiment, participants \revise{were} seated in a fixed chair and \revise{interacted} using a hand-held controller with a Ray Casting interface.
They \revise{initiated} each task by pressing a start button and \revise{were} allowed to take self-paced breaks between trials, including stretching or resting as needed.

\subsection{Participants}
\revise{A total of} 23 participants (9 males and 14 females) were recruited from the university’s participant recruitment system (SONA). The minimum required sample size was determined to be 22 using G*Power \cite{faul2009statistical} based on an effect size of 0.15, alpha level of 0.05, and power of 0.8 with the within-subject study design. Their average age is 19.35, with a range of 18 to 29 years. All participants have 20/20 (or corrected 20/20) vision, and they do not have physical impairments that would affect their use of VR devices.

Participants received 2.5 SONA credits in accordance with the university's SONA policy. According to the pre-questionnaire, 20 out of 23 participants had VR experience. These participants rated their familiarity with VR on a 7-point Likert scale \revise{(1 = Not at all, 7 = Very Familiar)}, reporting an average score of 3.2, \revise{indicating limited to moderate familiarity.}

\subsection{Apparatus}
The Vive Pro Head-Mounted Display (HMD) with a wireless adapter and a single controller is utilized. The HMD provides 110$^\circ$ vertical and horizontal FoV and 1440 $\times$ 1600 pixels display per eye (2880 $\times$ 1600 pixels combined). The experiment application is executed by Unity 2022.3.3f1 and on a Windows 10 desktop computer equipped with an Intel Xeon W-2245 CPU (3.90GHz), 64GB RAM, and an Nvidia GeForce RTX 3090 graphics card.
\section{Results}
To analyze detection time, accuracy, and timeout rate, we utilize a two-way Repeated Measures Analysis of Variance (RM-ANOVA) test with a significance level of $5\%$.
The NASA-TLX data is analyzed using a one-way RM-ANOVA. Following standard practice, we applied the Greenhouse-Geisser or Huynh-Feldt correction when Mauchly’s test indicated a violation of the sphericity assumption.
For significant interaction effects, we further conducted simple effect analyses to examine the effect of one factor at each level of the other factor.
For all significant main effects and simple effects, post-hoc comparisons were conducted using the Bonferroni correction to adjust for multiple comparisons, with $\alpha=.05$\revise{.}

\subsection{Detection Time}


\begin{table}
\includegraphics[angle=0, width=0.46\textwidth]{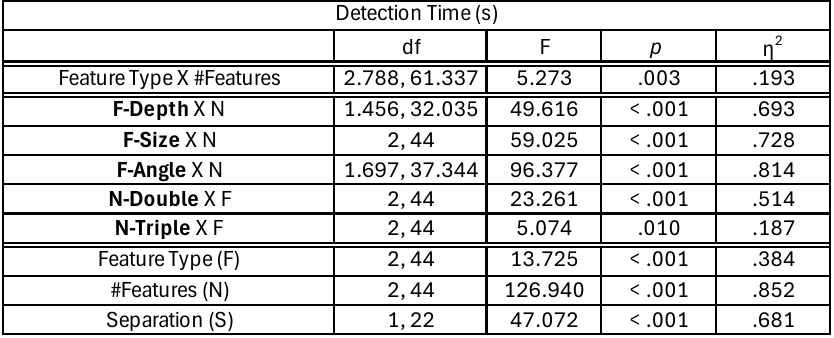}
\caption{
Detection time ANOVA results. One interaction effect and multiple main effects are found across conditions.
}
\label{table:detectionTime}
\end{table}

\begin{figure}[t]
\centering
\includegraphics[angle=0, width=\columnwidth]{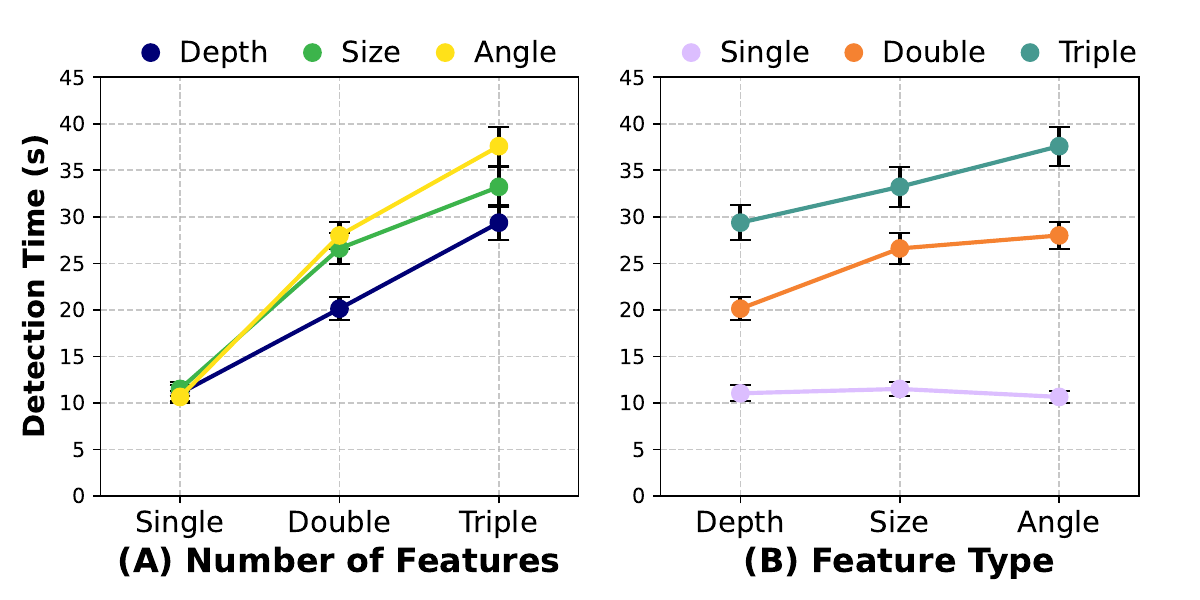}
\caption{
Detection time results show a significant interaction effect between Feature Type and Number of Features. Error bars represent standard errors of the mean.
The two graphs visualize the same interaction effect using different axis mappings\revise{.}
}
\label{figure:detectionTimeInteraction}
\end{figure}

\begin{figure}[t]
\includegraphics[angle=0, width=\columnwidth]{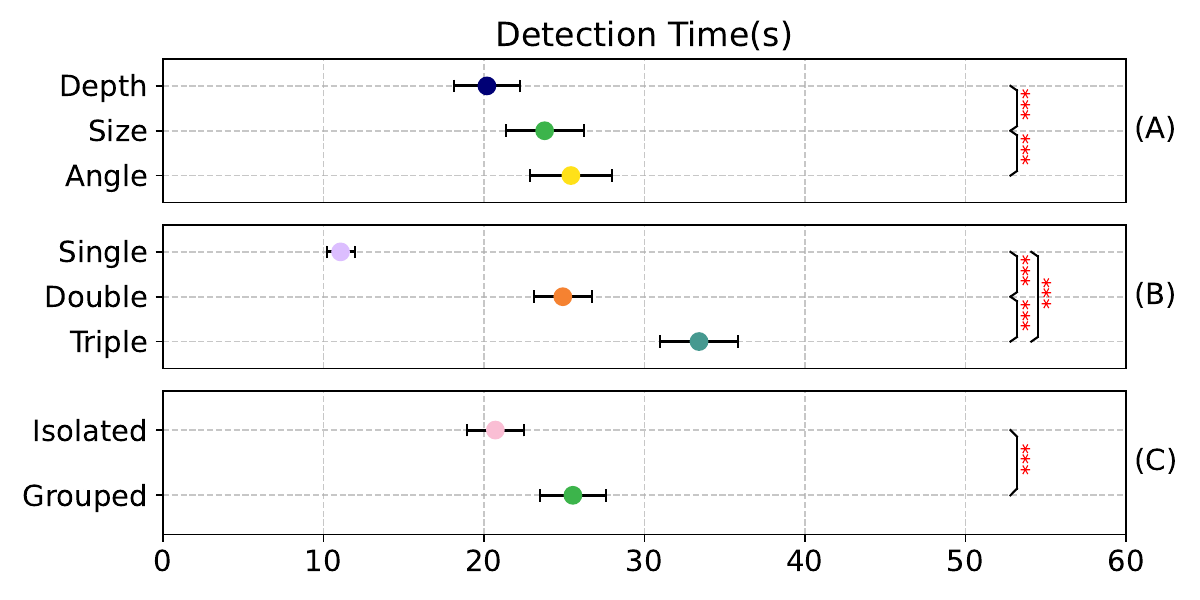}
\caption{
Detection time results with 95\% confidence interval. (A) Feature Type, (B) Number of Features, and (C) Separation. Pairwise comparisons of the main effects are indicated with brackets and asterisks: * ($p<.05$), ** ($p<.01$), and *** ($p<.001$).
}
\label{figure:detectionTime}
\end{figure}

Fig.~\ref{figure:detectionTime}, Fig.~\ref{figure:detectionTimeInteraction} and Table~\ref{table:detectionTime} present the ANOVA results of detection time.
A significant interaction effect is observed between the \textbf{Feature Type} and \textbf{Number of Features} variables ($p=.003$, Fig.~\ref{figure:detectionTimeInteraction}) 
Simple effect analyses reveal that within \textit{Depth} ($p<.001$), detection time significantly increased with \textbf{Number of Features}.
\textit{Single} ($M=11.0s$) has faster detection time than both \textit{Double} ($M=20.1s, p<.001$) and \textit{Triple} ($M=29.4s, p<.001$).
Detection time is also significantly faster in \textit{Double} than \textit{Triple} ($p<.001$).
Similarly, for \textbf{Size}, a simple effect of \textbf{Number of Features} is found ($p<.001$, Fig.~\ref{figure:detectionTimeInteraction}). \textit{Single} ($M=11.5s$) result in faster detection than  \textit{Double} ($M=26.6s, p<.001$) and \textit{Triple} ($M=33.2s, p<.001$), with \textit{Double} also faster than \textit{Triple} ($p=.007$).
For \textbf{Angle}, a simple effect of \textbf{Number of Features} is also observed  ($p<.001$, Fig.~\ref{figure:detectionTimeInteraction}).
\textit{Single} ($M=10.7s$) has a faster change detection time than \textit{Double} ($M=29.2s, p<.001$) and \textit{Triple} ($M=37.6s, p<.001$), with \textit{Double} faster than \textit{Triple} ($p=.002$).
Further, within \textit{Double}, a simple effect of \textbf{Feature Type} is found ($p<.001$).
\textit{Depth} ($M=20.1s$) enables faster detection than \textit{Size} ($M=26.6s, p<.001$) and \textit{Angle} ($M=29.2s, p<.001$).
There is also a simple effect of \textbf{Feature Type} within \textit{Triple} ($p=.010$).
\textit{Depth} ($M=29.4s$) leads faster detection than \textit{Angle} ($M=37.6s, p=.009$).

There is a main effect on \textbf{Feature Type} ($p<.001$, Fig.~\ref{figure:detectionTime}A). \textit{Depth} ($M=20.2s$) results in faster change detection than \textit{Size} ($M=23.8s, p=.007$) and \textit{Angle} ($M=25.8s, p<.001$).
A main effect is also found on \textbf{Number of Features} ($p<.001$, Fig.~\ref{figure:detectionTime}B).
\textit{Single} ($M=11.1s$) results in faster change detection compared to \textit{Double} ($M=25.3s, p<.001$) and \textit{Triple} ($M=33.4s, p<.001$).
\textit{Double} also results in faster change detection than \textit{Triple} ($p<.001$).
Furthermore, there is a main effect on \textbf{Separation} ($p<.001$, Fig.~\ref{figure:detectionTime}C).
\textit{Isolated} ($M=21.0s$) results in faster change detection than \textit{Grouped} ($M=25.5s, p<.001$).

\subsection{Timeout Rate}

\begin{table}[t]
\includegraphics[angle=0, width=0.46\textwidth]{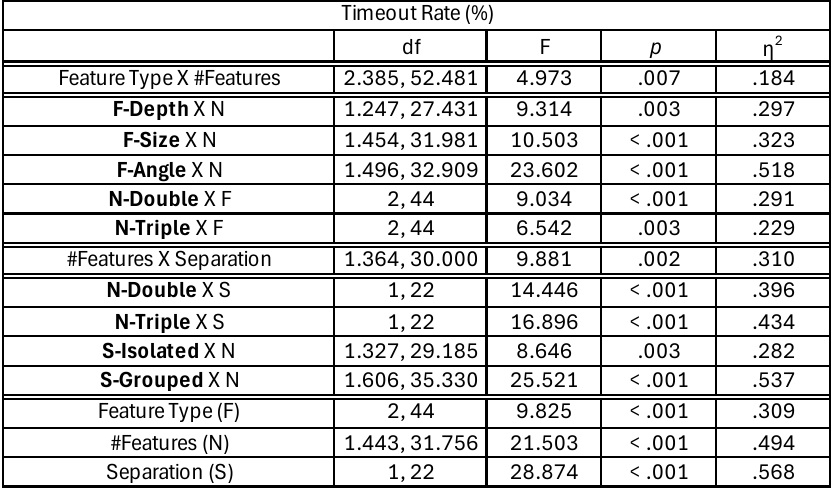}
\caption{
Timeout ANOVA results. Interaction effects and main effects are found across conditions.
}
\label{table:timeout}
\end{table}

\begin{figure}[t]
\centering
\includegraphics[angle=0, width=\columnwidth]{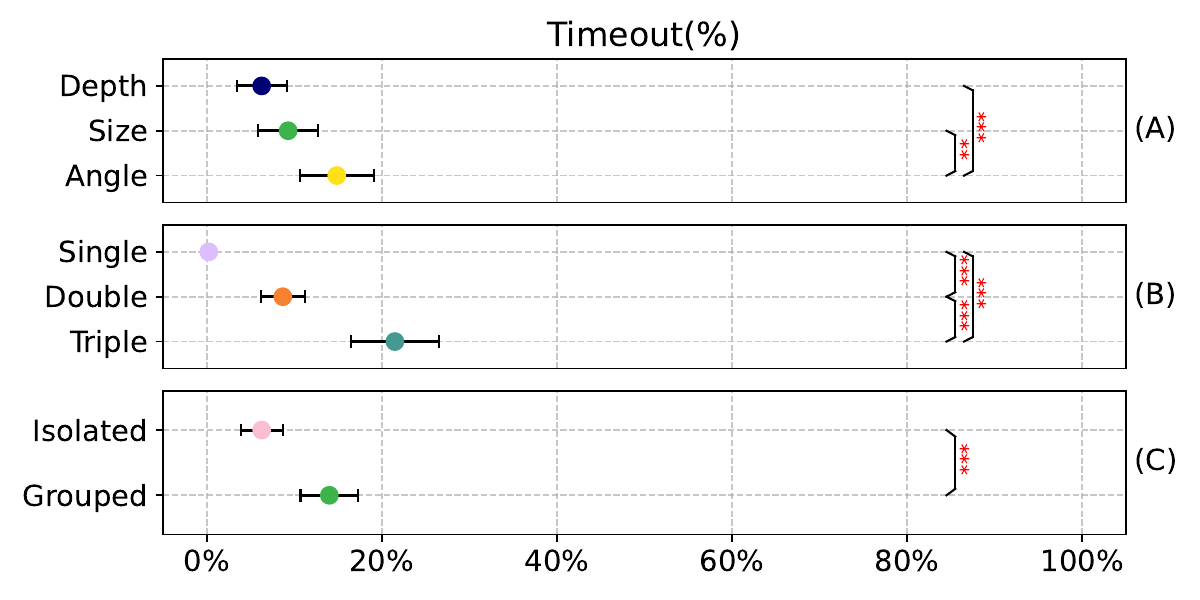}
\caption{
The timeout rate results in a 95\% confidence interval. (A) Feature Type, (B) Number of Features, and (C) Separation. Pairwise comparisons of the main effects are indicated with brackets and asterisks: * ($p<.05$), ** ($p<.01$), and *** ($p<.001$). }

\label{figure:timeout}
\end{figure}

\begin{figure}[t]
\centering
\includegraphics[angle=0, width=\columnwidth]{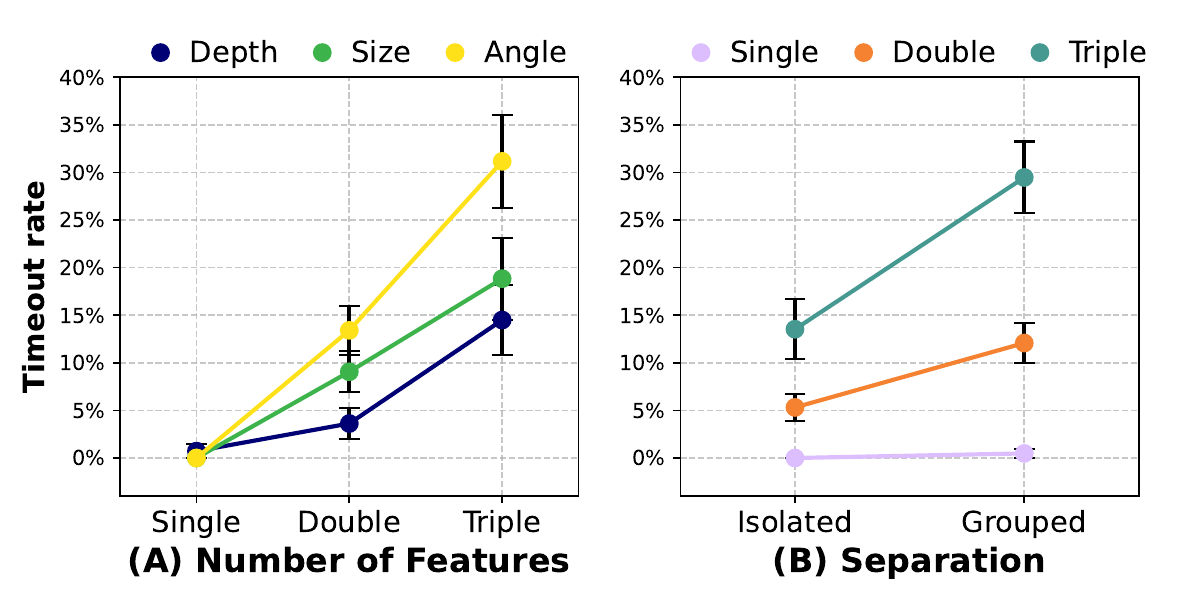}
\caption{
Timeout rate results show significant interaction effects (A) between Number of Features and Feature Type, and (B) between Separation and Number of Features. Error bars represent standard errors of the mean.
}
\label{figure:timeoutInteraction}
\end{figure}

Fig.~\ref{figure:timeout}, Fig.~\ref{figure:timeoutInteraction}, and Table~\ref{table:timeout} present the ANOVA result of the Timeout rate results.
There is an interaction effect between \textbf{Feature Type} and \textbf{Number of Features} ($p=.007$, Fig.~\ref{figure:timeoutInteraction}).
Simple effect analyses indicate that within \textit{Depth} ($p=.003$), timeout rates significantly increase with \textbf{Number of Features}.
\textit{Triple} ($M=14.5\%$) leads to a higher timeout rate than \textit{Single} ($M=0.7\%, p=.012$) and \textit{Double} ($M=3.6\%, p=.017$).
Similarly, a simple effect of \textbf{Number of Features} is found within \textit{Size} ($p<.001$, Fig.~\ref{figure:timeoutInteraction}).
\textit{Triple} has higher timeout rate ($M=18.8\%$) than \textit{Single} ($M=0.0\%, p=.004$), while \textit{Double} leads to a higher timeout rate ($M=9.1\%$) than \textit{Single} ($p=.010$).
For \textit{Angle}, a simple effect of \textbf{Number of Features} is also observed ($p<.001$).
\textit{Triple} has higher timeout rate ($M=31.2\%$) than \textit{Single} ($M=0.0\%, p<.001$) and \textit{Double} ($M=13.4\%, p=.005$), while \textit{Double} leads to a higher timeout rate than \textit{Single} ($p<.001$).
Additionally, a simple effect of \textbf{Feature Type} is found within \textit{Double} ($p<.001$).
\textit{Depth} ($M=3.6\%$) leads to a lower timeout rate than the \textit{Size} ($M=9.1\%, p=.046$) and \textit{Angle} ($M=13.4\%, p=.005$).
A simple effect of \textbf{Feature Type} is also found within \textit{Triple} ($p=.003$).
\textit{Angle} ($M=31.2\%$) has a higher timeout rate than \textit{Depth} ($M=14.5\%, p=.002$).

A significant interaction effect is also observed between the \textbf{Number of Features} and the \textbf{Separation} ($p=.002$, Fig.~\ref{figure:timeoutInteraction}).
A simple effect of \textbf{Separation} is found with \textit{Double} ($p<.001$).
\textit{Isolated} ($M=5.3\%$) leads to a lower timeout rate than \textit{Grouped} ($M=12.1\%, p<.001$).
There is also a simple effect of \textbf{Separation} with \textit{Triple} ($p<.001$). \textit{Isolated} ($M=13.5\%$) results in a lower timeout rate than \textit{Grouped} ($M=29.5\%, p<.001$).
Additionally, simple effect analyses reveal that within \textit{Isolated}, timeout rates increase significantly with \textbf{Number of Features} ($p=.003$).
\textit{Single} ($M=0.0\%$) results in a lower timeout rate than \textit{Double} ($M=5.3\%, p=.038$) and \textit{Triple} ($M=13.5\%, p=.010$).
Similarly, within \textit{Grouped}, a simple effect of \textbf{Number of Features} is found ($p<.001$).
\textit{Single} ($M=0.5\%$) leads to a lower timeout rate than \textit{Double} ($M=12.1\%, p<.001$) and \textit{Triple} ($M=29.5\%, p<.001$), with \textit{Double} being lower than \textit{Triple} ($p=.002$).

There is a main effect on \textbf{Feature Type} ($p<.001$, Fig. \ref{figure:timeout}A). \textit{Angle} ($M=14.9\%$) leads to a higher timeout rate than \textit{Size} ($M=9.3\%, p=.015$) and \textit{Depth} ($M=6.3\%, p<.001$). 
There is also a main effect on \textbf{Number of Features} ($p<.001$, Fig.~\ref{figure:timeout}B).
\textit{Single} ($M=0.2\%$) leads to a lower timeout rate than the \textit{Double} ($M=8.7$\%$, p=.002$) and \textit{Triple} ($M=21.5$\%$, p<.001$).
In addition, \textit{Double} leads to a lower timeout rate than \textit{Triple} ($p=.003$).
Finally, there is a main effect on \textbf{Separation} ($p<.001$, Fig.~\ref{figure:timeout}C).
\textit{Isolated} ($M=6.3\%$) leads to a lower timeout rate than \textit{Grouped} ($M=14.0\%, p<.001$).

\subsection{Accuracy}

\begin{table}
\includegraphics[angle=0, width=0.46\textwidth]{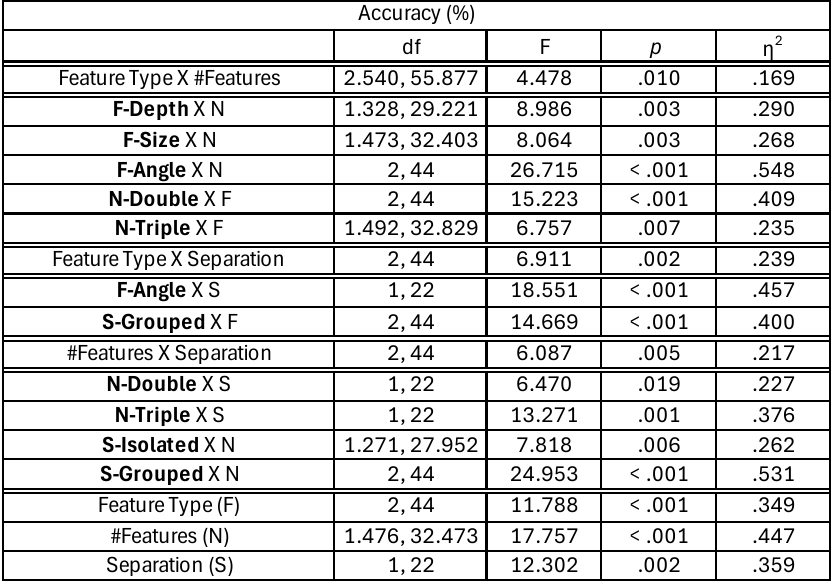}
\caption{
Accuracy ANOVA results. Interaction effects and main effects are found across conditions.
}
\vspace{-.5cm}
\label{table:accuracy}
\end{table}

\begin{figure}[t]
\centering
\includegraphics[angle=0, width=\columnwidth]{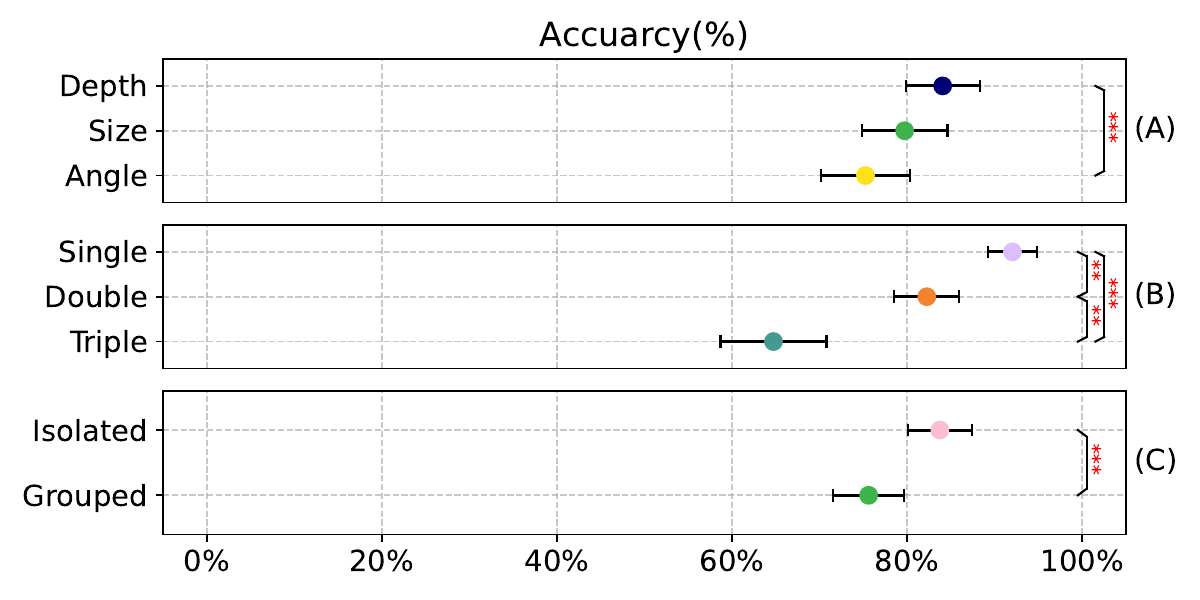}
\caption{
Accuracy results with 95\% confidence interval. (A) Feature Type, (B) Number of Features, and (C) Separation. Pairwise comparisons of the main effects are indicated with brackets and asterisks: * ($p<.05$), ** ($p<.01$),
and *** ($p<.001$).
}
\vspace{-.5cm}
\label{figure:accuracy}
\end{figure}

\begin{figure*}[t]
\centering
\includegraphics[angle=0, width=0.68\textwidth]{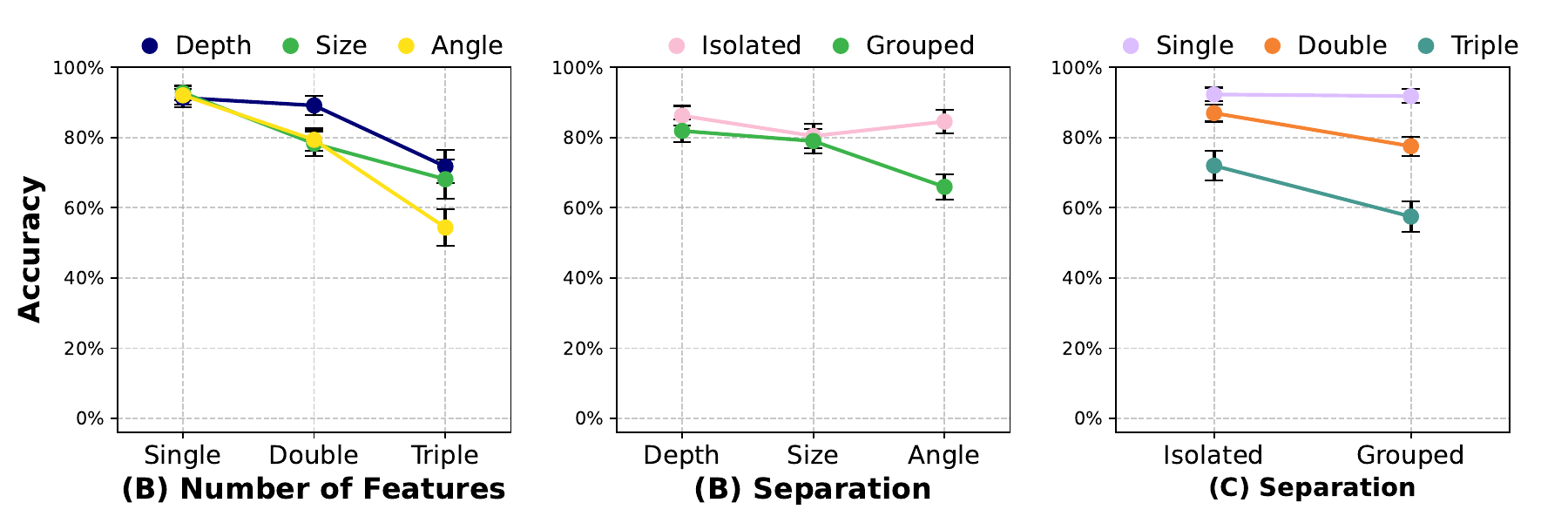}
\caption{
 Significant interaction effects on accuracy (A) between Number of Features and Feature Type, (B) between Feature Type and Separation, and (C) between Separation and Number of Features. Error bars represent standard errors of the mean.
}

\label{figure:accuracyInteraction}
\end{figure*}

Fig.~\ref{figure:accuracy}, Fig.~\ref{figure:accuracyInteraction}, and Table~\ref{table:accuracy} present the ANOVA results of the change detection accuracy.
There is an interaction effect between the \textbf{Feature Type} and \textbf{Number of Feature} ($p=.010$, Fig.~\ref{figure:accuracyInteraction}B).
Simple effect analyses within \textit{Depth} ($p=.003$), \textit{Triple} ($M=71.7\%$) shows lower accuracy than \textit{Single} ($M=91.3\%, p=.015$) and \textit{Double} ($M=89.1\%, p=.012$).
Within \textit{Size} ($p=.003$), show that \textit{Single} ($M=92.8\%$) results in higher accuracy than \textit{Double} ($M=78.3\%, p=.007$) and \textit{Triple} ($M=68.1\%, p=.011$).
Within \textit{Angle} ($p<.001$), \textit{Single} ($M=92.0\%$) results in higher accuracy than \textit{Double} ($M=75.4\%, p=.001$) and \textit{Triple} ($M=54.3\%, p<.001$).
\textit{Double} also has a higher accuracy than \textit{Triple} ($p=.003$).
Simple effect analyses reveal that, within \textit{Double} ($p<.001$), \textit{Depth} ($M=89.1\%$) results in higher accuracy than \textit{Size} ($M=78.3\%, p=.001$) and \textit{Angle} ($M=75.4\%, p<.001$).
Within \textit{Triple} ($p=.007$), \textit{Angle} ($M=54.3\%$) shows lower accuracy than \textit{Depth} ($M=71.7\%, p<.001$).

In addition, there is an interaction effect between \textbf{Feature Type} and \textbf{Separation} ($p=.002$, Fig.~\ref{figure:accuracyInteraction}C).
A simple effect is found within \textit{Angle} ($p<.001$), where \textit{Isolated} ($M=81.9\%$) results in higher accuracy than \textit{Grouped} ($M=65.9\%, p<.001$).
Also, within \textit{Grouped} ($p<.001$), accuracy is lower for \textit{Angle} ($M=65.9\%$) than for \textit{Size} ($M=79.0\%, p=.005$) and \textit{Depth} ($M=81.9\%, p<.001$).

There is an interaction effect between the \textbf{Number of Features} and \textbf{Separation} ($p=.005$, Fig.~\ref{figure:accuracyInteraction}A).
Simple effect analyses reveal that, within \textit{Double} ($p=.019$), accuracy is higher for \textit{Isolated} ($M=84.3\%$) than \textit{Grouped} ($M=77.5\%$, $p=.019$).
Similarly with \textit{Triple} ($p=.001$), \textit{Isolated} ($M=72.0\%$) show higher higher accuracy than \textit{Grouped} ($M=57.5\%$, $p=.001$).
A simple effect of \textbf{Number of Features} with \textit{Isolated} ($p=.006$), \textit{Single} ($M=92.3\%$) result in higher accuracy than \textit{Double} ($M=84.3\%$, $p=.019$) and \textit{Triple} ($M=72.0\%$, $p=.013$).
A simple effect of \textbf{Number of Features} within \textit{Grouped} ($p<.001$), \textit{Single} ($M=91.8\%$) shows higher accuracy than \textit{Double} ($M=77.5\%$, $p=.006$) and \textit{Triple} ($M=57.5\%$, $p<.001$).
\textit{Double} also has a higher accuracy than \textit{Triple} ($p=.003$).

A main effect is found on \textbf{Feature Type} ($p<.001$, Fig.~\ref{figure:accuracy}A).
\textit{Depth} ($M=84.1\%, p<.001$) and \textit{Size} ($M=79.7\%, p=.041$) result in higher accuracy than \textit{Angle} ($M=73.9\%$).
In addition, a main effect of \textbf{Number of Features} ($p<.001$, Fig.~\ref{figure:accuracy}B) shows that \textit{Single} ($M=92.0\%$) yields higher accuracy than \textit{Double} ($M=80.9\%, p=.003$) and \textit{Triple} ($M=64.7\%, p<.001$), with \textit{Double} also higher than \textit{Triple} ($p=.011$).
Finally, there is a main effect on \textbf{Separation} ($p=.002$, Fig.~\ref{figure:accuracy}C). \textit{Isolated} ($M=82.9\%$) has a higher accuracy than \textit{Grouped} ($M=75.6\%, p=.002$).

\subsection{NASA-TLX}


\dhkim{The ANOVA results for the NASA-TLX questionnaire responses showed significant main effects across all subscales of the questionnaire.
(Mental Demand: $F(2, 44)= 39.885$, $p<.001$, $\eta^2_p=.645$; Physical Demand: $F(2, 44)=6.198$, $p=.004$, $\eta^2_p=.220$; Temporal Demand: $F(2, 44)=7.507$, $p=.002$, $\eta^2_p=.254$; Performance: $F(2, 44)=17.149$, $p<.001$, $\eta^2_p=.438$; Effort: $F(1.599, 35.173)=10.894$, $p<.001$, $\eta^2_p=.331$; Frustration: $F(1.608, 35.376)=9.026$, $p=.001$, $\eta^2_p=.291$)
}

The \textit{Single} condition resulted in significantly lower Mental Demand ($M=2.48, p<.001$), Effort ($M=3.61, p=.045$), and Frustration ($M=1.70, p=.005$), as well as better perceived Performance ($M=5.87, p=.001$), compared to the \textit{Double} condition (Mental Demand: $M=4.04$; Effort: $M=4.70$; Frustration: $M=2.70$; Performance: $M=4.65$).

Compared to the \textit{Triple} condition (Mental Demand: $M=4.83$; Physical Demand: $M=1.65$; Temporal Demand: $M=3.61$, Effort: $M=5.13$, Frustration: $M=3.26$, Performance: $M=3.70$), the \textit{Single} condition also showed significantly lower Mental Demand ($p<.001$), Physical Demand ($M=1.17, p=.002$), and Temporal Demand ($M=2.57, p<.001$), as well as lower Effort ($p<.001$) and Frustration ($p< .001$), and better Performance ($p<.001$).

When comparing the \textit{Double} and \textit{Triple} conditions, the \textit{Double} condition only showed significantly lower Mental Demand ($p=.032$), while no significant differences were observed in other NASA-TLX dimensions.

\subsection{Result Summary}
The results consistently support our hypotheses (\textbf{H1–H4}) across detection time, timeout rate, and accuracy. 
 
Supporting \textbf{H1}, the \textit{Depth} condition consistently resulted in faster detection time (\textbf{H1-1}), lower timeout rates (\textbf{H1-2}), and higher accuracy (\textbf{H1-3}) compared to the \textit{Size} and \textit{Angle} conditions.
The interaction effects between \textbf{Number of Features} and \textbf{Feature Type} indicated that the performance advantage of \textit{Depth} was relatively preserved even as the number of features increased.

Consistent with \textbf{H2}, the \textit{Single} condition showed the best performance, with faster detection time (\textbf{H2-1}), lower timeout rates (\textbf{H2-2}), and higher accuracy (\textbf{H2-3}) compared to both the \textit{Double} and \textit{Triple} conditions. Similarly, the \textit{Double} condition outperformed the \textit{Triple} condition in detection time (\textbf{H2-4}), timeout rate (\textbf{H2-5}), and accuracy (\textbf{H2-6}). These results confirm that increasing the number of visual features disrupted efficient detection and required more effortful cognitive processing.
Interaction effects between \textbf{Number of Features} and \textbf{Feature Type} were observed in detection time, timeout rate, and accuracy.
In particular, the performance degradation from increasing features was most severe for the \textit{Angle} condition, while the \textit{Depth} condition was more robust to feature complexity.
\dhkim{
In contrast, the degradation of the \textit{Size} condition was severe when the feature complexity increased from \textit{Single} to \textit{Double}. However, when it increased from \textit{Double} to \textit{Triple}, the performance decline was less severe than in the \textit{Angle} condition.
}

Supporting \textbf{H3}, the \textit{Isolated} condition resulted in faster detection time (\textbf{H3-1}), lower timeout rates (\textbf{H3-2}), and higher accuracy (\textbf{H3-3}) than the \textit{Grouped} condition, confirming that spatial separation facilitated efficient change detection, likely by reducing distraction from nearby objects. Interestingly, interaction effects between \textbf{Separation} and \textbf{Number of Features} were observed for both timeout rate and accuracy. Specifically, the performance benefit of the \textit{Isolated} condition became more pronounced as the number of features increased, particularly in the \textit{Double} and \textit{Triple} conditions.

Our results also mostly support \textbf{H4}, which predicted that increasing the number of visual features would increase perceived workload. Specifically, the \textit{Single} condition resulted in the lowest workload across most NASA-TLX dimensions, including lower mental demand, effort, and frustration, and better perceived performance compared to both the \textit{Double} and \textit{Triple} conditions, supporting \textbf{H4-1}.
Additionally, the \textit{Double} condition showed lower mental demand and better performance than the \textit{Triple} condition, supporting \textbf{H4-2}. However, there were no significant differences in physical demand and temporal demand between the \textit{Single} and \textit{Double} conditions, nor between the \textit{Double} and \textit{Triple} conditions. This suggests that increasing the number of features did not substantially affect participants' perceived physical or time-related demands when the complexity increase was moderate. Nevertheless, a significant difference was observed between the \textit{Single} and \textit{Triple} conditions, indicating that a larger increase in feature complexity was required to impact participants’ perception of physical effort and time pressure.

We also examined head-turning count as a potential behavioral measure of engagement or attention. However, no statistically significant differences were found across conditions, and we therefore omitted it from further analysis.
\section{Discussion}

Grounded in our results, we discuss key findings that address \textbf{RQ1} to \textbf{RQ3} and offer possible explanations for observed results.


\subsection{Preattentive Features Support Efficient Change Detection: Depth as the Most Robust Feature}

 In response to \textbf{RQ1}, we found that all features (\textit{Depth}, \textit{Size}, and \textit{Angle}) were equally effective in simple scenes (\textit{Single} condition), showing no significant performance differences. 
This finding suggests that, in visually simple environments, features such as size, orientation, and depth are equally salient and can be efficiently processed without requiring extensive attentional resources. However, as visual complexity increased (\textit{Double} and \textit{Triple} feature conditions), the influence of feature type became more pronounced. \textit{Depth} changes remained highly detectable even in complex scenes, while \textit{Angle} changes consistently resulted in the poorest performance across all conditions. \textit{Size} changes exhibited intermediate performance, adapting depending on the level of scene complexity.

A possible explanation for the superior performance of \textit{Depth} changes is that depth perception engages specialized spatial processing mechanisms that remain robust even under visual complexity. According to the Visuospatial Sketchpad model~\cite{logie2014visuo}, spatial properties such as depth are processed by the inner scribe subsystem, which manages spatial and movement-related information and is less susceptible to overload compared to the visual cache, which stores intrinsic object features like size or orientation.
In addition, depth cue integration principles suggest that the visual system combines multiple depth cues (e.g., stereopsis, motion parallax, occlusion, and relative size), weighting them by reliability ~\cite{cutting1995perceiving, surdick1997perception, howard2012perceiving}.
Immersive VR provides particularly strong support for these cues through stereoscopic rendering and head-tracking, making depth signals more reliable and salient than size or orientation, which depend on more ambiguous perceptual information. 
These rich spatial signals enhance the salience of depth changes, allowing them to produce strong figure-ground separation and ``pop-out'' effects, even in visually cluttered scenes.
\dhkim{
In contrast, detecting \textit{Angle} changes requires finer perceptual discrimination and relies more heavily on feature-based processing, making them more vulnerable to interference as visual complexity increases.
}

\dhkim{
One interesting finding is the flexible role of size as a perceptual cue, which appears to shift based on the user's cognitive load.
Size information can serve two different roles in a 3D environment: one is an object's scale, and the other is depth as a monocular cue. Our results indicate that, depending on the overall scene complexity, the perceptual system may prioritize one of these functions over the other.
This interpretation is supported by the observed trend in detection time and accuracy. In the \textit{Size} condition, there was a sharp performance decline from \textit{Single} to \textit{Double}, but the drop was less pronounced than in the \textit{Angle} condition when complexity increased to \textit{Triple}.
This suggests that as the scene becomes more cluttered and the main visual channels are overloaded, the human perceptual system relies more heavily on the size's depth cue role.
This functional shift, from treating size as a feature to be identified to using it as a cue for spatial localization, explains its mitigated performance decline under high visual load and has important implications for designers using size to encode information in complex VR scenes.
}

\subsection{Visual Complexity Impairs Change Detection Performance and Increases Perceived Workload}


Addressing \textbf{RQ2}, our findings show that increasing visual complexity, operationalized by the number of attentive features, negatively affected change detection performance and increased perceived workload in VR. Increasing visual complexity made it more difficult for participants to quickly and accurately detect changes, resulting in longer detection times and higher timeout rates. As feature complexity increased, competing visual information within the scene likely interfered with participants' ability to detect and localize changes accurately.

Participants also reported significantly higher perceived workload, particularly in terms of mental demand, effort, and frustration. These findings indicate that increasing visual complexity not only challenges visual search but also makes the detection task more cognitively demanding. Notably, perceived physical and temporal demand remained stable across the \textit{Single}, \textit{Double}, and \textit{Triple} conditions, with a noticeable increase only observed between the \textit{Single} and \textit{Triple} conditions. This suggests that higher feature complexity was required to impact physical or time-related workload.

\dhkim{
This trend can be explained by the target's reduced visual distinctiveness in more complex scenes. In the simple scene (Single condition), the target's unique feature made it stand out, allowing for effortless and instantaneous detection. The visual system could identify this noticeable change without individually inspecting each object.
In contrast, when multiple features varied, no single property made the target unique.
Locating the change thus required a conscious, effortful search where participants had to mentally combine multiple features to identify the target.
This cognitively taxing task of binding and comparing features naturally led to the slower detection times, increased errors, and higher subjective workload we observed, as noted in prior visualization research~\cite{munzner2014visualization}.
}

 \subsection{Spatial Separation Facilitates Change Detection Under Visual Complexity}
Addressing \textbf{RQ3}, our results demonstrate that spatial separation significantly improved change detection performance in VR, particularly under higher levels of visual complexity. Across all conditions, the \textit{Isolated} spatial arrangement consistently resulted in faster detection time, higher accuracy, and lower timeout rates compared to the \textit{Grouped} condition. Notably, this performance advantage became more pronounced as visual complexity increased.

A possible explanation for this effect is that spatial separation reduces visual clutter and minimizes competition between objects for attentional resources. When a target object is spatially isolated, users can more easily direct their attention to it without scanning through densely populated regions. In contrast, when objects are grouped, particularly under high complexity, users must search more extensively and compare multiple similar objects, increasing task difficulty.
Moreover, the decline in change detection performance observed in the \textit{Grouped} condition may be explained by perceptual grouping that influences how visual information is encoded and stored. The human visual system tends to integrate spatially proximate objects with similar features into a single perceptual unit, which can hinder the encoding of individual object features. 

This interpretation relates to a long-standing debate in VSTM research regarding whether multi-attribute objects are stored as integrated units or as separate features. While one perspective suggests that VSTM capacity is determined by the number of whole objects stored~\cite{vogel2001storage}, other research indicates that VSTM capacity is also influenced by the number and complexity of features within objects~\cite{alvarez2004capacity}. 
Our results support this latter view, suggesting VSTM capacity is affected by both the number of objects and their feature complexity.
In the context of our experiment, where participants were required to detect changes in specific object attributes (e.g., depth, size, angle) under varying levels of visual complexity, our findings provide supporting evidence that perceptual grouping may interfere with feature-based encoding, particularly in visually cluttered scenes.

\subsection{Design Considerations for VR Systems}
Our findings highlight several principles relevant to both VR system design and immersive analytics applications. Depth cues consistently supported faster and more accurate detection than size or orientation, suggesting that depth should be emphasized when encoding critical updates or anomalies. At the same time, excessive feature complexity impaired performance and increased workload, underscoring the need to minimize clutter and carefully select which visual features convey key information. Spatial isolation further improved detection, indicating that separating task-relevant objects or data subsets from surrounding elements can reduce interference and support efficient recognition.

\textbf{Implications for Immersive Analytics.} These findings provide concrete guidance for immersive analytics systems that support data exploration in VR and AR \cite{han2023evaluating, han2025perception}. For immersive analytics systems, these tendencies imply that depth encodings could be particularly effective for guiding attention in large 3D data spaces, while orientation changes are less reliable and size cues are context dependent. By prioritizing depth, controlling visual complexity, and leveraging spatial separation, designers can reduce cognitive load and better support rapid sensemaking in immersive environments.

\subsection{Limitation and Future Work}
While our study examined the effects of preattentive features on change detection, we acknowledge that the task context does not strictly reflect classical preattentive processing conditions. In traditional definitions, preattentive processing occurs rapidly, automatically, and often within very brief exposure times, without the need for active search or attention shifts \cite{ware2019information, munzner2014visualization}. In contrast, our experimental design required participants to turn their heads and search for changes in a dynamic VR environment, which likely engaged both preattentive and attentive processing mechanisms. Therefore, while our study has shown that the selected visual features (size, angle, depth) and spatial separation can effectively support visual perception for change detection, their effects in our study should be interpreted with caution as facilitating early-stage perception rather than reflecting purely preattentive processing. Future work is needed to isolate these processes further or to examine them in more controlled environments.

\revise{
The participant pool primarily consisted of university students. While age was reported, racial and ethnic demographic data were not collected, which limits the assessment of population diversity. In addition, the sample reflects a relatively narrow age range, normal or corrected-to-normal vision, and a university-educated population, and therefore is not representative of the general population. However, because the study employed a within-subject design, the primary analyses are less sensitive to between-participant demographic variability in perceptual performance. Although there is no established evidence that race directly affects low-level visual perception, demographic factors may correlate with differences in technology familiarity and interaction strategies in VR, which could influence task performance.}

Moreover, while we measured detection time as a key performance metric, we did not exclude the duration associated with participants turning their heads. This is because determining the precise onset and offset of head movement is challenging in dynamic VR settings. Although we posit that head-turning duration is unlikely to have significantly impacted the detection time results, we acknowledge this limitation and the associated uncertainty. Future studies could incorporate finer-grained motion tracking or alternative experimental designs to disentangle the contributions of head movement from visual detection performance.
\section{Conclusion}
In this study, we investigated how preattentive features influence change detection performance in immersive virtual environments with varying levels of visual complexity.
The results show that preattentive features support rapid and accurate detection in simple scenes but become less effective as visual complexity increases.
Specifically, depth changes were detected faster and more reliably than size or angle changes, particularly in complex conditions, suggesting that spatial information plays a more dominant role in guiding attention in VR.
Furthermore, our results show that the \revise{size} change can be perceived in different roles, either as an actual size or as a depth cue, according to the complexity of the scene.
These findings contribute to a better understanding of how preattentive feature type, scene complexity, and spatial separation influence change detection, offering valuable insights for the design of more effective immersive environments.


\bibliographystyle{abbrv-doi-hyperref}

\bibliography{reference}

@book{howard2012perceiving,
  title={Perceiving in depth, volume 1: basic mechanisms},
  author={Howard, Ian P},
  year={2012},
  publisher={Oxford University Press}
}

@article{surdick1997perception,
  title={The perception of distance in simulated visual displays: A comparison of the effectiveness and accuracy of multiple depth cues across viewing distances},
  author={Surdick, R Troy and Davis, Elizabeth T and King, Robert A and Hodges, Larry F},
  journal={Presence: Teleoperators \& Virtual Environments},
  volume={6},
  number={5},
  pages={513--531},
  year={1997},
  publisher={MIT Press 238 Main St., Suite 500, Cambridge, MA 02142-1046, USA journals~…}
}

@incollection{cutting1995perceiving,
  title={Perceiving layout and knowing distances: The integration, relative potency, and contextual use of different information about depth},
  author={Cutting, James E and Vishton, Peter M},
  booktitle={Perception of space and motion},
  pages={69--117},
  year={1995},
  publisher={Elsevier}
}

@article{han2025perception,
  title={Perception of Visual Variables on Virtual Wall-Sized Tiled Displays in Immersive Environments},
  author={Han, Dongyun and Bezerianos, Anastasia and Isenberg, Petra and Cho, Isaac},
  journal={IEEE Transactions on Visualization and Computer Graphics},
  year={2025},
  publisher={IEEE}
}

@inproceedings{han2023evaluating,
  title={Evaluating 3D User Interaction Techniques on Spatial Working Memory for 3D Scatter Plot Exploration in Immersive Analytics},
  author={Han, Dongyun and Cho, Isaac},
   booktitle={IEEE International Symposium on Mixed and Augmented Reality (ISMAR), 2023.},

}

@book{munzner2014visualization,
  title={Visualization analysis and design},
  author={Munzner, Tamara},
  year={2014},
  publisher={CRC press}
}

@article{peterson2013gestalt,
  title={The Gestalt principle of similarity benefits visual working memory},
  author={Peterson, Dwight J and Berryhill, Marian E},
  journal={Psychonomic bulletin \& review},
  volume={20},
  pages={1282--1289},
  year={2013},
  publisher={Springer}
}

@article{cowan2008differences,
  title={What are the differences between long-term, short-term, and working memory?},
  author={Cowan, Nelson},
  journal={Progress in brain research},
  volume={169},
  pages={323--338},
  year={2008},
  publisher={Elsevier}
}

@article{cowan2005capacity,
  title={On the capacity of attention: Its estimation and its role in working memory and cognitive aptitudes},
  author={Cowan, Nelson and Elliott, Emily M and Saults, J Scott and Morey, Candice C and Mattox, Sam and Hismjatullina, Anna and Conway, Andrew RA},
  journal={Cognitive psychology},
  volume={51},
  number={1},
  pages={42--100},
  year={2005},
  publisher={Elsevier}
}

@article{moriya2019visual,
  title={Visual-working-memory training improves both quantity and quality},
  author={Moriya, Jun},
  journal={Journal of Cognitive Enhancement},
  volume={3},
  number={2},
  pages={221--232},
  year={2019},
  publisher={Springer}
}

@article{simons1997change,
  title={Change blindness},
  author={Simons, Daniel J and Levin, Daniel T},
  journal={Trends in cognitive sciences},
  volume={1},
  number={7},
  pages={261--267},
  year={1997},
  publisher={Elsevier}
}

@article{simons2005change,
  title={Change blindness: Theory and consequences},
  author={Simons, Daniel J and Ambinder, Michael S},
  journal={Current directions in psychological science},
  volume={14},
  number={1},
  pages={44--48},
  year={2005},
  publisher={SAGE Publications Sage CA: Los Angeles, CA}
}

@article{alvarez2004capacity,
  title={The capacity of visual short-term memory is set both by visual information load and by number of objects},
  author={Alvarez, George A and Cavanagh, Patrick},
  journal={Psychological science},
  volume={15},
  number={2},
  pages={106--111},
  year={2004},
  publisher={SAGE Publications Sage CA: Los Angeles, CA}
}

@article{awh2007visual,
  title={Visual working memory represents a fixed number of items regardless of complexity},
  author={Awh, Edward and Barton, Brian and Vogel, Edward K},
  journal={Psychological science},
  volume={18},
  number={7},
  pages={622--628},
  year={2007},
  publisher={SAGE Publications Sage CA: Los Angeles, CA}
}

@article{xu2006dissociable,
  title={Dissociable neural mechanisms supporting visual short-term memory for objects},
  author={Xu, Yaoda and Chun, Marvin M},
  journal={Nature},
  volume={440},
  number={7080},
  pages={91--95},
  year={2006},
  publisher={Nature Publishing Group UK London}
}

@incollection{hart1988development,
  title={Development of NASA-TLX (Task Load Index): Results of empirical and theoretical research},
  author={Hart, Sandra G and Staveland, Lowell E},
  booktitle={Advances in psychology},
  volume={52},
  pages={139--183},
  year={1988},
  publisher={Elsevier}
}

@inproceedings{kim2023comparative,
  title={Comparative analysis of change blindness in virtual reality and augmented reality environments},
  author={Kim, Donghoon and Han, Dongyun and Cho, Isaac},
  booktitle={2023 IEEE International Symposium on Mixed and Augmented Reality (ISMAR)},
  pages={990--998},
  year={2023},
  organization={IEEE}
}

@article{martin2023study,
  title={A study of change blindness in immersive environments},
  author={Martin, Daniel and Sun, Xin and Gutierrez, Diego and Masia, Belen},
  journal={IEEE Transactions on Visualization and Computer Graphics},
  volume={29},
  number={5},
  pages={2446--2455},
  year={2023},
  publisher={IEEE}
}

@inproceedings{steinicke2010change,
  title={Change blindness phenomena for stereoscopic projection systems},
  author={Steinicke, Frank and Bruder, Gerd and Hinrichs, Klaus and Willemsen, Pete},
  booktitle={2010 IEEE Virtual Reality Conference (VR)},
  pages={187--194},
  year={2010},
  organization={IEEE}
}

@article{simons1998failure,
  title={Failure to detect changes to people during a real-world interaction},
  author={Simons, Daniel J and Levin, Daniel T},
  journal={Psychonomic Bulletin \& Review},
  volume={5},
  pages={644--649},
  year={1998},
  publisher={Springer}
}

@article{levin2002memory,
  title={Memory for centrally attended changing objects in an incidental real-world change detection paradigm},
  author={Levin, Daniel T and Simons, Daniel J and Angelone, Bonnie L and Chabris, Christopher F},
  journal={British Journal of Psychology},
  volume={93},
  number={3},
  pages={289--302},
  year={2002},
  publisher={Wiley Online Library}
}

@article{varakin2007comparison,
  title={Comparison and representation failures both cause real-world change blindness},
  author={Varakin, D Alexander and Levin, Daniel T and Collins, Krista M},
  journal={Perception},
  volume={36},
  number={5},
  pages={737--749},
  year={2007},
  publisher={SAGE Publications Sage UK: London, England}
}

@article{wolfe2019preattentive,
  title={What is a preattentive feature?},
  author={Wolfe, Jeremy M and Utochkin, Igor S},
  journal={Current opinion in psychology},
  volume={29},
  pages={19--26},
  year={2019},
  publisher={Elsevier}
}

@book{ware2019information,
  title={Information visualization: perception for design},
  author={Ware, Colin},
  year={2019},
  publisher={Morgan Kaufmann}
}

@article{treisman1988feature,
  title={Feature analysis in early vision: evidence from search asymmetries.},
  author={Treisman, Anne and Gormican, Stephen},
  journal={Psychological review},
  volume={95},
  number={1},
  pages={15},
  year={1988},
  publisher={American Psychological Association}
}

@article{nakayama1986serial,
  title={Serial and parallel processing of visual feature conjunctions},
  author={Nakayama, Ken and Silverman, Gerald H},
  journal={Nature},
  volume={320},
  number={6059},
  pages={264--265},
  year={1986},
  publisher={Nature Publishing Group UK London}
}

@article{treisman1985preattentive,
  title={Preattentive processing in vision},
  author={Treisman, Anne},
  journal={Computer vision, graphics, and image processing},
  volume={31},
  number={2},
  pages={156--177},
  year={1985},
  publisher={Elsevier}
}

@article{healey2011attention,
  title={Attention and visual memory in visualization and computer graphics},
  author={Healey, Christopher and Enns, James},
  journal={IEEE transactions on visualization and computer graphics},
  volume={18},
  number={7},
  pages={1170--1188},
  year={2011},
  publisher={IEEE}
}

@article{wolfe1992role,
  title={The role of categorization in visual search for orientation.},
  author={Wolfe, Jeremy M and Friedman-Hill, Stacia R and Stewart, Marion I and O'Connell, Kathleen M},
  journal={Journal of Experimental Psychology: Human Perception and Performance},
  volume={18},
  number={1},
  pages={34},
  year={1992},
  publisher={American Psychological Association}
}

@article{treisman1980feature,
  title={A feature-integration theory of attention},
  author={Treisman, Anne M and Gelade, Garry},
  journal={Cognitive psychology},
  volume={12},
  number={1},
  pages={97--136},
  year={1980},
  publisher={Elsevier}
}

@article{cheal1992attention,
  title={Attention in visual search: Multiple search classes},
  author={Cheal, Marylou and Lyon, Don R},
  journal={Perception \& Psychophysics},
  volume={52},
  number={2},
  pages={113--138},
  year={1992},
  publisher={Springer}
}

@article{nagy1990visual,
  title={Visual search for color differences with foveal and peripheral vision},
  author={Nagy, Allen L and Sanchez, Robert R and Hughes, Thomas C},
  journal={JOSA A},
  volume={7},
  number={10},
  pages={1995--2001},
  year={1990},
  publisher={Optica Publishing Group}
}

@article{healey1996high,
  title={High-speed visual estimation using preattentive processing},
  author={Healey, Christopher G and Booth, Kellogg S and Enns, James T},
  journal={ACM Transactions on Computer-Human Interaction (TOCHI)},
  volume={3},
  number={2},
  pages={107--135},
  year={1996},
  publisher={ACM New York, NY, USA}
}

@article{treisman1982perceptual,
  title={Perceptual grouping and attention in visual search for features and for objects.},
  author={Treisman, Anne},
  journal={Journal of experimental psychology: human perception and performance},
  volume={8},
  number={2},
  pages={194},
  year={1982},
  publisher={American Psychological Association}
}

@article{enns1990three,
  title={Three-dimensional features that pop out in visual search.},
  author={Enns, James T},
  year={1990},
  publisher={Taylor \& Francis}
}

@article{wolfe1989guided,
  title={Guided search: an alternative to the feature integration model for visual search.},
  author={Wolfe, Jeremy M and Cave, Kyle R and Franzel, Susan L},
  journal={Journal of Experimental Psychology: Human perception and performance},
  volume={15},
  number={3},
  pages={419},
  year={1989},
  publisher={American Psychological Association}
}

@article{xu2007visual,
  title={Visual short-term memory benefit for objects on different 3-D surfaces.},
  author={Xu, Yaoda and Nakayama, Ken},
  journal={Journal of Experimental Psychology: General},
  volume={136},
  number={4},
  pages={653},
  year={2007},
  publisher={American Psychological Association}
}

@article{viswanathan2002dynamics,
  title={Dynamics of attention in depth: Evidence from multi-element tracking},
  author={Viswanathan, Lavanya and Mingolla, Ennio},
  journal={Perception},
  volume={31},
  number={12},
  pages={1415--1437},
  year={2002},
  publisher={SAGE Publications Sage UK: London, England}
}

@article{sagehorn2024comparative,
  title={A comparative analysis of face and object perception in 2D laboratory and virtual reality settings: insights from induced oscillatory responses},
  author={Sagehorn, Merle and Kisker, Joanna and Johnsdorf, Marike and Gruber, Thomas and Sch{\"o}ne, Benjamin},
  journal={Experimental Brain Research},
  pages={1--19},
  year={2024},
  publisher={Springer}
}

@article{barrera2023preattentive,
  title={How the Preattentive process is exploited in practical information visualization design: A review},
  author={Barrera-Leon, Luisa and Corno, Fulvio and De Russis, Luigi},
  journal={International Journal of Human--Computer Interaction},
  volume={39},
  number={4},
  pages={707--720},
  year={2023},
  publisher={Taylor \& Francis}
}

@article{snow2014real,
  title={Real-world objects are more memorable than photographs of objects},
  author={Snow, Jacqueline C and Skiba, Rafal M and Coleman, Taylor L and Berryhill, Marian E},
  journal={Frontiers in human neuroscience},
  volume={8},
  pages={837},
  year={2014},
  publisher={Frontiers Media SA}
}

@article{korisky2021dimensions,
  title={Dimensions of perception: 3D real-life objects are more readily detected than their 2D images},
  author={Korisky, Uri and Mudrik, Liad},
  journal={Psychological Science},
  volume={32},
  number={10},
  pages={1636--1648},
  year={2021},
  publisher={SAGE Publications Sage CA: Los Angeles, CA}
}

@article{mullen2021time,
  title={Time compression in virtual reality},
  author={Mullen, Grayson and Davidenko, Nicolas},
  journal={Timing \& Time Perception},
  volume={9},
  number={4},
  pages={377--392},
  year={2021},
  publisher={Brill}
}

@article{bansal2019movement,
  title={Movement-contingent time flow in virtual reality causes temporal recalibration},
  author={Bansal, Ambika and Weech, S{\'e}amas and Barnett-Cowan, Michael},
  journal={Scientific reports},
  volume={9},
  number={1},
  pages={4378},
  year={2019},
  publisher={Nature Publishing Group UK London}
}

@article{rzepka2023familiar,
  title={Familiar size affects perception differently in virtual reality and the real world},
  author={Rzepka, Anna M and Hussey, Kieran J and Maltz, Margaret V and Babin, Karsten and Wilcox, Laurie M and Culham, Jody C},
  journal={Philosophical Transactions of the Royal Society B},
  volume={378},
  number={1869},
  pages={20210464},
  year={2023},
  publisher={The Royal Society}
}

@article{plumert2005distance,
  title={Distance perception in real and virtual environments},
  author={Plumert, Jodie M and Kearney, Joseph K and Cremer, James F and Recker, Kara},
  journal={ACM Transactions on Applied Perception (TAP)},
  volume={2},
  number={3},
  pages={216--233},
  year={2005},
  publisher={ACM New York, NY, USA}
}

@inproceedings{cho2012evaluating,
  title={Evaluating depth perception of volumetric data in semi-immersive VR},
  author={Cho, Isaac and Dou, Wenwen and Wartell, Zachary and Ribarsky, William and Wang, Xiaoyu},
  booktitle={Proceedings of the international working conference on advanced visual interfaces},
  pages={266--269},
  year={2012}
}

@article{bowman2007virtual,
  title={Virtual reality: how much immersion is enough?},
  author={Bowman, Doug A and McMahan, Ryan P},
  journal={Computer},
  volume={40},
  number={7},
  pages={36--43},
  year={2007},
  publisher={IEEE}
}

@article{creem2005influence,
  title={The influence of restricted viewing conditions on egocentric distance perception: Implications for real and virtual indoor environments},
  author={Creem-Regehr, Sarah H and Willemsen, Peter and Gooch, Amy A and Thompson, William B},
  journal={Perception},
  volume={34},
  number={2},
  pages={191--204},
  year={2005},
  publisher={SAGE Publications Sage UK: London, England}
}

@inproceedings{drascic1996perceptual,
  title={Perceptual issues in augmented reality},
  author={Drascic, David and Milgram, Paul},
  booktitle={Stereoscopic displays and virtual reality systems III},
  volume={2653},
  pages={123--134},
  year={1996},
  organization={Spie}
}

@article{vienne2020depth,
  title={Depth perception in virtual reality systems: effect of screen distance, environment richness and display factors},
  author={Vienne, Cyril and Masfrand, St{\'e}phane and Bourdin, Christophe and Vercher, Jean-Louis},
  journal={IEEE Access},
  volume={8},
  pages={29099--29110},
  year={2020},
  publisher={IEEE}
}

@inproceedings{hart2006nasa,
  title={NASA-task load index (NASA-TLX); 20 years later},
  author={Hart, Sandra G},
  booktitle={Proceedings of the human factors and ergonomics society annual meeting},
  volume={50},
  number={9},
  pages={904--908},
  year={2006},
  organization={Sage publications Sage CA: Los Angeles, CA}
}

@article{roth2017visual,
  title={Visual variables},
  author={Roth, Robert E},
  journal={International encyclopedia of geography: People, the earth, environment and technology},
  pages={1--11},
  year={2017},
  publisher={John Wiley \& Sons, Ltd}
}

@article{buss2018visual,
  title={Visual working memory in early development: a developmental cognitive neuroscience perspective},
  author={Buss, Aaron T and Ross-Sheehy, Shannon and Reynolds, Greg D},
  journal={Journal of Neurophysiology},
  volume={120},
  number={4},
  pages={1472--1483},
  year={2018},
  publisher={American Physiological Society Bethesda, MD}
}

@article{hulleman2020medium,
  title={Medium versus difficult visual search: How a quantitative change in the functional visual field leads to a qualitative difference in performance},
  author={Hulleman, Johan and Lund, Kristofer and Skarratt, Paul A},
  journal={Attention, Perception, \& Psychophysics},
  volume={82},
  pages={118--139},
  year={2020},
  publisher={Springer}
}

@article{wolfe2020forty,
  title={Forty years after feature integration theory: An introduction to the special issue in honor of the contributions of Anne Treisman},
  author={Wolfe, Jeremy M},
  journal={Attention, Perception, \& Psychophysics},
  volume={82},
  pages={1--6},
  year={2020},
  publisher={Springer}
}

@book{logie2014visuo,
  title={Visuo-spatial working memory},
  author={Logie, Robert H},
  year={2014},
  publisher={Psychology Press}
}

@article{vogel2001storage,
  title={Storage of features, conjunctions, and objects in visual working memory.},
  author={Vogel, Edward K and Woodman, Geoffrey F and Luck, Steven J},
  journal={Journal of experimental psychology: human perception and performance},
  volume={27},
  number={1},
  pages={92},
  year={2001},
  publisher={American Psychological Association}
}

@article{h1997inner,
  title={The inner eye and the inner scribe of visuo-spatial working memory: Evidence from developmental fractionation},
  author={H. Logie, Robert and Pearson, David G},
  journal={European Journal of cognitive psychology},
  volume={9},
  number={3},
  pages={241--257},
  year={1997},
  publisher={Taylor \& Francis}
}

@article{faul2009statistical,
  title={Statistical power analyses using G* Power 3.1: Tests for correlation and regression analyses},
  author={Faul, Franz and Erdfelder, Edgar and Buchner, Axel and Lang, Albert-Georg},
  journal={Behavior research methods},
  volume={41},
  number={4},
  pages={1149--1160},
  year={2009},
  publisher={Springer}
}

@article{krekhov2019deadeye,
  title={Deadeye visualization revisited: Investigation of preattentiveness and applicability in virtual environments},
  author={Krekhov, Andrey and Cmentowski, Sebastian and Waschk, Andre and Kr{\"u}ger, Jens},
  journal={IEEE transactions on visualization and computer graphics},
  volume={26},
  number={1},
  pages={547--557},
  year={2019},
  publisher={IEEE}
}

@article{hadnett2019effect,
  title={The effect of task on visual attention in interactive virtual environments},
  author={Hadnett-Hunter, Jacob and Nicolaou, George and O'Neill, Eamonn and Proulx, Michael},
  journal={ACM Transactions on Applied Perception (TAP)},
  volume={16},
  number={3},
  pages={1--17},
  year={2019},
  publisher={ACM New York, NY, USA}
}

@inproceedings{barreiros2016pre,
  title={Pre-attentive features in natural augmented reality visualizations},
  author={Barreiros, Carla and Veas, Eduardo and Pammer-Schindler, Viktoria},
  booktitle={2016 IEEE international symposium on mixed and augmented reality (ISMAR-Adjunct)},
  pages={72--73},
  year={2016},
  organization={IEEE}
}

@article{steinicke2011change,
  title={Change blindness phenomena for virtual reality display systems},
  author={Steinicke, Frank and Bruder, Gerd and Hinrichs, Klaus and Willemsen, Pete},
  journal={IEEE Transactions on visualization and computer graphics},
  volume={17},
  number={9},
  pages={1223--1233},
  year={2011},
  publisher={IEEE}
}

@article{senel2023imperceptible,
  title={Imperceptible body transformation in virtual reality: Saliency of self representation},
  author={Senel, Gizem and Macia-Varela, Francisco and Gallego, Jaime and Jensen, Hatice Pehlivan and Hornb{\ae}k, Kasper and Slater, Mel},
  journal={Iscience},
  volume={26},
  number={10},
  year={2023},
  publisher={Elsevier}
}

\end{document}